\documentclass[aps,prl,twocolumn,reprint,superscriptaddress,nofootinbib,
tightenlines,preprintnumbers,showkeys,floatfix]{revtex4-1}

\usepackage[utf8]{inputenc}
\usepackage[english]{babel}
\usepackage{amssymb,amsthm,amsmath,amstext,amsbsy,amsopn,mathrsfs}
\usepackage{bbm}
\usepackage{nicefrac}
\usepackage{slashed}
\usepackage{graphicx}
\usepackage{hyperref}
\usepackage{leftidx}
\usepackage{environ}
\usepackage{mathtools}
\usepackage{xspace}
\usepackage{array}
\usepackage{xcolor}
\usepackage{pgfplotstable}
\usepackage{isotope}
\usepackage{suffix}
\usepackage{booktabs,colortbl}

\newcommand{\ie}{\textit{i.e.}\xspace}

\newcommand{\apriori}{\textit{a priori}\xspace}
\WithSuffix\newcommand\apriori*{\textit{a-priori}\xspace}

\newcommand{\mathspace}{\ \ }
\newcommand{\mathtext}[1]{\mathspace\text{#1}\mathspace}

\newcommand{\vecr}{\mathbf{r}}
\newcommand{\vecx}{\mathbf{x}}

\newcommand{\vecn}{\mathbf{n}}

\newcommand{\vZero}{\mathbf{0}}

\newcommand{\vdelta}{\delta^{(3)}}

\newcommand{\dd}{\mathrm{d}}

\newcommand{\ii}{\mathrm{i}}
\newcommand{\ee}{\mathrm{e}}

\newcommand{\OO}{\mathcal{O}}
\newcommand{\ZZ}{\mathbb{Z}}
\newcommand{\RR}{\mathbb{R}}

\newcommand{\bra}[1]{\langle #1|}

\newcommand{\ket}[1]{|#1\rangle}

\newcommand{\braket}[2]{\langle #1|#2\rangle}

\newcommand{\mbraket}[3]{\langle #1|#2|#3\rangle}

\newcommand{\abs}[1]{\left|#1\right|}

\newcommand*\rvec[1]%
{\ensuremath{\overset{\smash{\raisebox{-1.5pt}{\tiny$\rightarrow$}}}{#1}}}
\newcommand*\lvec[1]%
{\ensuremath{\overset{\smash{\raisebox{-1.5pt}{\tiny$\leftarrow$}}}{#1}}}

\NewEnviron{subalign}[1][]{%
\begin{subequations}\begin{align}
  \BODY
\end{align}\label{#1}\end{subequations}
}

\NewEnviron{spliteq}{%
\begin{equation}\begin{split}
  \BODY
\end{split}\end{equation}
}

\NewEnviron{mleq}{%
\begin{equation}
  \BODY
\end{equation}
}

\newcommand{\psiinf}{\psi_{\infty}}

\newcommand{\Vd}{\Delta V}

\renewcommand{\vec}[1]{\mathbf{#1}}

\newcommand{\psip}{\tilde\psi}
\newcommand{\Edp}{\Delta\tilde{E}}
\newcommand{\EdL}{\Delta E^*}

\newcommand{\TopRule}{\toprule[1.25pt]}
\newcommand{\BottomRule}{\TopRule}
\newcommand{\MidRule}{\midrule[0.5pt]}
\newcommand{\MidDoubleRule}{\hline\hline}

\begin{document}

\title{Charged-particle bound states in periodic boxes}

\author{Hang Yu}
\email{yhang@ncsu.edu}
\affiliation{Department of Physics, North Carolina State University,
Raleigh, NC 27695, USA}

\author{Sebastian König}
\email{skoenig@ncsu.edu}
\affiliation{Department of Physics, North Carolina State University,
Raleigh, NC 27695, USA}

\author{Dean Lee}
\email{leed@frib.msu.edu}
\affiliation{Facility for Rare Isotope Beams \& Department of Physics
and Astronomy, Michigan State University, MI 48824, USA}

\begin{abstract}
We consider the binding energy of a two-body system with a repulsive Coulomb
interaction in a finite periodic volume.
We define the finite-volume Coulomb potential as the usual Coulomb potential,
except that the distance is defined as the shortest separation between the two
bodies in the periodic volume.
We investigate this problem in one and three-dimensional periodic boxes and
derive the asymptotic behavior of the volume dependence for bound states with
zero angular momentum in terms of Whittaker functions.
We benchmark our results against numerical calculations and show how the method
can be used to extract asymptotic normalization coefficients for
charged-particle bound states.
The results we derive here have immediate applications for calculations of
atomic nuclei in finite periodic volumes for the case where the leading
finite-volume correction is associated with two charged clusters.
\end{abstract}

\maketitle

Finite-volume (FV) simulations in cubic boxes with periodic boundary conditions
have emerged as a well established theoretical technique to study quantum
systems.
Pioneered by Lüscher in a series of highly influential
papers~\cite{Luscher:1985dn,Luscher:1986pf,Luscher:1990ux} that showed how
real-world properties of a quantum system are encoded in how its discrete energy
levels change as the volume size is varied, the method has become a
standard approach, for example, in Lattice Quantum Chromodynamics (LQCD) to
extract scattering information for hadronic systems.
Over the past decade, driven by progress in computing that enables
simulations of increasingly complex systems, extending the technique
in various directions has become an area of very active
research~\cite{Kreuzer:2010ti,Kreuzer:2012sr,Polejaeva:2012ut,Briceno:2012rv,%
Kreuzer:2013oya,Meissner:2014dea,Hansen:2015zga,Hammer:2017uqm,%
Hammer:2017kms,Mai:2017bge,Doring:2018xxx,Pang:2019dfe,Culver:2019vvu,%
Briceno:2019muc,Romero-Lopez:2019qrt,Hansen:2020zhy,Muller:2021uur}.
Moreover, few-body approaches formulated in FV can be used to match and
extrapolate LQCD results to an effective field theory (EFT)
description~\cite{Barnea:2013uqa,Detmold:2021oro,Kirscher:2015yda,%
Yaron:2022rmb}.

When simulated in FV, bound-state energy levels have an exponential
dependence on the size $L$ of the periodic box that encodes asymptotic
properties of the state's wave function in infinite
volume~\cite{Luscher:1985dn,Konig:2011nz,Konig:2011ti}.
For a general bound state of $N\geq2$ particles with lowest breakup
into two clusters with $A$ and $N-A$ particles, respectively, the volume
dependence of the (binding) energy, $\Delta E_N(L) \equiv E_N(\infty) - E_N(L)$,
with $E_N(L) = {-}B_N(L)$ denoting the energy at volume $L$,
has been shown to be~\cite{Konig:2017krd,Konig:2020lzo}
\begin{multline}
 \Delta E_N(L) = (-1)^{\ell+1} \sqrt{\tfrac{2}{\pi}} f(d)
  \,\mu_{A|N-A}^{{-}1}\abs{A_\infty}^2 \\
 \null \times \kappa^{2-d/2}_{A|N-A}L^{1-d/2}
 K_{d/2-1}(\kappa_{A|N-A}L) \,,
\label{eq:ANC-FV}
\end{multline}
with $d$ the number spatial dimensions, $f(d)$ a normalization
factor, and $K_{d/2-1}$ a modified Bessel function.
$\kappa_{A|N-A} = \sqrt{\mathstrut2\mu_{A|N-A}(B_{N}-B_{A}-B_{N-A})}$
with the reduced mass $\mu_{A|N-A}$ of the two-cluster system
and the cluster binding energies $B_{A}$ and $B_{N-A}$ is the
relevant momentum scale.
Moreover, $A_\infty$ is the asymptotic normalization coefficient (ANC) of the
cluster wave function, a quantity that plays an important role for the
description of low-energy capture processes.
Equation~\eqref{eq:ANC-FV} implies that both $\kappa_{A|N-A}$ and $A_\infty$ can
be extracted by fitting the volume dependence of numerical simulations.
This is in principle an efficient way to compute ANCs from \textit{ab initio}
calculations at finite volume.
Unfortunately, most cases of astrophysical interest involve two clusters that
each have nonzero electric charge, and the analogous expression for
Eq.~\eqref{eq:ANC-FV} for charged clusters had not been derived.

In this Letter we address this problem and determine the leading volume
dependence for bound states composed of two charged particles.
While there have been studies of finite-volume electromagnetic corrections using
perturbation theory~\cite{%
Davoudi:2014qua,Beane:2014qha,Borsanyi:2014jba,Stellin:2020gst,Guo:2021qfu,%
Guo:2021lhz}, we derive the analytic form with fully nonperturbative Coulomb
repulsion.
The nonperturbative nature of the Coulomb interaction becomes important in
medium-mass and heavy nuclei with a significant number of protons.
The work presented here has immediate applications to finite-volume simulations
of bound nuclei using \textit{ab initio} Lattice EFT~\cite{%
Lee:2008fa,Lahde:2019npb,Lu:2021tab,Shen:2022bak,Elhatisari:2022qfr}, and the
definition of
the finite-volume Coulomb interaction we use here is the same as in
Lattice EFT.
We derive the analog for Eq.~\eqref{eq:ANC-FV} when the relevant continuum
threshold corresponds to two charged clusters.
While we focus on the case of two charged point particles, our results can be
regarded as the leading term for the multipole expansions of the permanent
and/or induced cluster charge densities.

\paragraph{Derivation}
We consider a two-body system of particles interacting via a finite-range
central potential $V$ plus a repulsive Coulomb potential $V_C$.
For simplicity we assume that $V$ is local,
$\mbraket{\vecr}{V}{\vecr'} = V(r) \, \vdelta(\vecr-\vecr')$,
but all results remain valid for a general nonlocal short-range potential.
The interaction range $R$ is the smallest distance for which it holds that
$V(r)=0$ if $r = \abs{\vecr} > R$.
The Coulomb potential is given by
\begin{equation}
 V_C(r) = \frac{\gamma}{2\mu r}
 \mathtext{,} \gamma = 2\mu \alpha Z_1 Z_2 > 0 \,,
\label{eq:V-C}
\end{equation}
where $\mu$ denotes the reduced mass of the two-body system, $\alpha \approx
1/137$ is the electromagnetic fine-structure constant, and $Z_{1,2}$ are the
charges of the two particles.
The Hamiltonian of the system is given by $H = H_0 + V + V_C$
with the kinetic-energy operator $H_0$.
If we consider the system enclosed in a periodic cubic box with edge length $L$,
the Hamiltonian becomes
\begin{equation}
 H_L = H_0 + V_{\{L\}} + V_{C,\{L\}} \,.
\label{eq:H-L}
\end{equation}
The finite-range potential is easily made periodic by defining
$V_{\{L\}}(\vecx) = \sum_{\vecn} V(\vecx - \vecn L)$.
Such a definition remains valid up to negligible corrections for short-range
potentials that do not have a strict finite range but fall off faster than any
power law.
For the Coulomb potential, however, the long-range tail $\sim r^{{-}1}$
complicates matters.
To obtain a well-defined periodic extension of $V_C$, we let $V_{C,\vec{n}L}$ be
a shifted version of $V_C$, centered at $\vec{n}L$ and clipped such that it is
nonzero only within the box of edge length $L$ around its center.
With $d$ the dimension of our space and $B=[-L/2,L/2)$, we define
$V_{C,\vec{n}L} = \theta_{B^d}(\vecr-\vecn L) V_C(\vecr-\vecn L)$ with
$\theta_{B^d}(\vecr) = 1$ for $\vecr \in B^d$ and vanishing otherwise.
Equipped with these clipped and shifted potentials, we can now define
\begin{equation}
 V_{C,\{L\}}(\vecr) = \sum_{\vecn} V_{C,\vecn L}(\vecr) \,.
\end{equation}
Effectively, this
definition implies that we simply let the Coulomb tail grow with the box.
We note that Lattice QCD+QED calculations use a different approach and define
a periodic FV Coulomb potential $U_L(\vecr)$ by subtracting the ``zero mode''
in momentum space~\cite{Beane:2014qha}.
In the Supplemental Material~\cite{SuppMat} we show that this $U_L(\vecr)$ can
be expanded to have exactly $V_{C,\{L\}}(\vecr)$ as first term, with a dominant
correction contributing an $\OO(1/L)$ constant shift to the binding energy.

We assume now that the total potential $V+V_C$ is such that the system supports
an $S$-wave bound state $\ket{\psiinf}$ with energy
${-}E_\infty<0 = {-}\kappa_\infty^2/(2\mu)$ in infinite-volume, \ie,
$H\ket{\psi_\infty} = (H_0 + V_C + V)\ket{\psi_\infty}
= {-}E_\infty\ket{\psi_\infty}$,
and our goal is to derive an expression for the finite-volume
energy shift $\Delta E(L) = E_\infty - E(L)$, where $E(L)$ denotes the energy of
the state at volume $L$, \ie, $H_L\ket{\psi_L} = E(L)\ket{\psi_L}$.
In a simplified setup that considers the Coulomb potential~\eqref{eq:V-C} in one
spatial dimension, it is possible to directly use the boundary condition
imposed on the finite-volume wave function at the edge of the box to
obtain the energy shift as
\begin{equation}
 \Delta E(L)
 = \frac{\kappa_\infty A_\infty^2}{\mu} \ee^{\ii\pi\bar{\eta}}
 \frac{W'_{{-}\bar{\eta},\frac12}(\kappa L)}
 {W'_{\bar{\eta},\frac12}({-}\kappa L)}
 + \OO\left[\ee^{{-}2\kappa L}\right] \,,
 \label{eq:DeltaE-1D-final}
\end{equation}
where $A_\infty$ is the ANC of the
infinite-volume wave function, $W_{\bar{\eta},\frac12}(z)$ with
$\bar{\eta} = \gamma/(2\kappa_\infty)$ is a Whittaker function, and the
prime denotes the derivative with respect to the argument.
We point out that phase factor in
Eq.~\eqref{eq:DeltaE-1D-final} is crucial to ensure that overall $\Delta E(L)$
is real.
A full derivation of this result is provided in the Supplemental
Material~\cite{SuppMat}.

For the three-dimensional system, the infinite-volume bound-state wave function
for $r = \abs{\vecx} \geq R$ is given by
$\psiinf(\vecx)
= A_\infty {W_{{-}\bar\eta,1/2}(2\kappa r)}/{(\sqrt{4\pi} r)}$.
A simple treatment of the 3D system based on the periodic boundary condition is
not possible because the long-range tail $~ r^{-1}$ combined with the breaking
of spherical symmetry by the box-shaped finite volume renders the setup too
complicated.
We therefore use an alternative formalism, starting with
an intermediate Hamiltonian $\tilde{H}_{L} = H_0 + V_{C,\{L\}} + V$
that includes the truncated periodic Coulomb potential, but leaves the
short-range potential $V$ as in infinite volume.
We can write
$\tilde{H}_{L} = H + \epsilon \Vd_C \mathtext{for} \epsilon\to1$,
with
\begin{align}
 &\Vd_C = V_{C,\{L\}} - V_C
 = \sum_{\vecn\in\ZZ^3}\Vd_{C,(\vec 0,\vecn)L} \,, \\
 &\Vd_{C,(\vecn,\vecn')L}(\vecx) = \theta_{B^3}(\vecx - \vecn'L) \nonumber \\
 &\hspace{8em}\null\times\left[
  V_{C,\vecn'L}(\vecx) - V_C(\vecx-\vecn L)
 \right] \,.
\label{eq:def_shiftc3D}
\end{align}
For the exact ground state $\ket{\psip_L}$ of $\tilde{H}_L$ we have
$\tilde{H}_L\ket{\psip_L} = {-}\tilde{E}(L)\ket{\psip_L}$,
and we can now treat $\Vd_C$ as a perturbation on top of this state.
To that end we write
\begin{multline}
 (H+\epsilon\Vd_C)(\ket{\psip_L^{(0)}}+\epsilon \ket{\psip_L^{(1)}} + \cdots) \\
 = (\tilde{E}^{(0)} + \epsilon \Delta\tilde{E}^{(1)}+\cdots)
 (\ket{\psip_L^{(0)}}+\epsilon \ket{\psip_L^{(1)}} + \cdots) \,,
\label{eq:psi-L-PT-exp}
\end{multline}
where to lowest order we have
$\ket{\psip^{(0)}} = \ket{\psi_\infty}$ and $\tilde{E}^{(0)} = {-}E_\infty$.
The leading volume dependence will be found below in terms of just
$\ket{\psip^{(0)}}$, and the main purpose of the perturbative treatment is that
it allows us to derive explicit bounds for subleading corrections.
To conclude the basic setup, we note that at this stage all $L$ dependence
comes from the definition of $\Vd_C$, and there is no $L$-periodic boundary
condition imposed on $\ket{\psip_L}$.
We use $\ket{\psip_L}$ in order to construct an ansatz
for the actual ground state $\ket{\psi_L}$ of $H_L$.
In the following we write equations explicitly in configuration space
and define
\begin{equation}
 \psip_{L,0}(\vec x) = \sum_{\vecn\in\ZZ^3} \psip_L(\vec x-\vec n L) \,,
 \label{eq:psip-L-0}
\end{equation}
\ie, we approximate the exact finite-volume solution by summing shifted
copies of the exact eigenstate of $\tilde{H}_{L}$ introduced above.
Using the ansatz $\psip_{L,0}(\vec x)$, we can follow steps very similar to
the derivation for neutral particles~\cite{Konig:2011nz,Luscher:1985dn}.
Applying $H_L$ to $\psip_{L,0}(\vec x)$, we obtain
\begin{multline}
 H_L \psip_{L,0}(\vec{x})
 = {-}\tilde{E}(L) \psip_{L,0}(\vec{x})
 + \!\! \sum_{\vec{n},\vec{n}'\neq \vec{n}} \!\!
 V(\vec{x} -\vec{n}'L) \\
 \null \times \psip_L(\vec{x} - \vec{n}L)
 \equiv {-}\tilde{E}(L)\psip_{L,0}(\vec{x}) + \zeta(\vecx) \,,
\label{eq:2ndstepp}
\end{multline}
with
$\zeta(\vecx) = \sum_{\vec{n}}\sum_{\vec{n}'\neq \vec{n}}
V(\vec{x} -\vec{n}'L)\psip_L(\vec{x} - \vec{n}L)$.
This function only involves the short-range interaction $V$ and
scales as $\zeta(\vecx)\sim\OO(\ee^{{-}\kappa L})$ for large $\abs{\vecx}$.
For the exact finite-volume solution $\ket{\psi_L}$, it holds that
$\ket{\psi_L} = \beta\ket{\psip_{L,0}} + \ket{\psip'_L}$,
with $\ket{\tilde{\psi}'_L} = \OO(\ee^{{-}\kappa L})$ and $\beta$ chosen
so that $\braket{\psip'_L}{\psip^{\phantom'}_{L,0}} = 0$.
We emphasize here that if instead we had used the naive
ansatz $\psi_{L,0}(\vec x)$ that replaces $\psip_L$ with $\psiinf$ in
Eq.~\eqref{eq:psip-L-0},
as it is appropriate for neutral particles, we would end up with the
weaker asymptotic scaling $\OO(\ee^{{-}\kappa L/2})$ for $\ket{\tilde{\psi}_L}$,
which would lead to unacceptably large subleading corrections.
We now write the overall energy shift at volume $L$ as
$\Delta E(L) = \EdL(L) + \Edp(L)$,
where $\Edp(L) = E_\infty - \tilde{E}(L)$ and
\begin{equation}
 \EdL(L) = E(L) - \tilde{E}(L) =
 \frac{\braket{\psip_{L,0}}{\zeta}}{\braket{\psip_{L,0}}{\psip_{L,0}}}
 + \OO(\ee^{-3\kappa L/2}) \,.
\label{eq:PT-error1}
\end{equation}
The leading volume dependence stems from $\EdL(L)$, for which with
\begin{equation}
 \zeta(\vecx) = \sum_{|\vec n|=1} V(\vecx)\psip_L(\vecx-\vecn L)
 + \OO(\ee^{{-}\sqrt{2}\kappa L}) \,,
\end{equation}
and $\braket{\psip_{L,0}}{\psip_{L,0}} = 1 + \OO(\ee^{{-}\kappa L})$ we obtain
\begin{equation}
 \EdL(L) = \sum_{|\vec n|=1} \int_{B^3} \dd^3 x \,
 \psip_L(\vec x) V(\vec x)\psip_L(\vecx-\vecn L)
\label{eq:EdL-0}
\end{equation}
up to terms $\OO(\ee^{{-}\sqrt{2}\kappa L})$.
This expression can be further evaluated by considering explicitly
$\vec n = {-}\hat z$ and exploiting the symmetry of the wave function.
As we explain further in the Supplemental Material~\cite{SuppMat}, this
ultimately leads to
\begin{multline}
 \Delta E(L) = \underbrace{%
  {-}\frac{3A_\infty^2}{\mu L}
  \Big[W_{-\bar\eta,\frac{1}{2}}(\kappa L)\Big]^2
 }_{\equiv \Delta E_0(L)} + \Edp(L) + \Edp'(L) \\[-1em]
 \null + \OO(\ee^{{-}\sqrt{2}\kappa L}) \,.
\label{eq:DeltaE-3D-final}
\end{multline}
Besides higher-order exponential corrections---which actually involve
Whittaker functions but can be expressed as $\OO(\ee^{{-}\sqrt{2}\kappa L})$
asymptotically---there are two correction terms in
Eq.~\eqref{eq:DeltaE-3D-final}: $\Edp(L)$ has been defined already in
above Eq.~\eqref{eq:PT-error1}, and $\Edp'(L)$ is explained further in the
Supplemental Material~\cite{SuppMat}.
Our perturbative setup makes it possible to derive the asymptotic scaling of
both these correction terms, the full details of which are also presented in
the Supplemental Material~\cite{SuppMat}.
It turns out that although their detailed forms are quite different, both terms
actually scale the same asymptotically, namely
\begin{equation}
 \Edp(L), \Edp'(L) =
 \OO\!\left(\frac{\bar\eta}{(\kappa L)^2}\right) \times \Delta E_{0}(L) \,,
\end{equation}
where $\Delta E_{0}(L)$ is the leading volume dependence as defined in
Eq.~\eqref{eq:DeltaE-3D-final}.

\paragraph{Numerical examples}
We evaluate the performance of Eq.~\eqref{eq:DeltaE-3D-final} with explicit
numerical calculations.
To that end we use the ``generator code'' of Ref.~\cite{Konig:2017krd},
extended to include the Coulomb interaction, to compute
1D and 3D bound states in boxes with a range of sizes.
To the volume dependence of these energies we then fit the appropriate
expressions for $\Delta E(L)$ to extract the infinite-volume binding momenta and
ANCs.
We use units that set $\hbar = 1$ and express
all physical quantities in terms of the particle mass
$m$, which we also set to unity.

The generator code constructs the finite-volume Hamiltonian based on a
simple lattice discretization (see Ref.~\cite{Konig:2020lzo} for details).
Because $V_C(r)$ is singular at $r=0$, we regularize it at short
distances.
While the simplest way to achieve this is a simple cut (setting $V_C(r) = 0$
for $r$ less than some small range),
we instead multiply $V_C(r)$ by a Gauss regulator $(1-\ee^{{-}r^2/R_C^2})$.
Such a smooth regulator is preferable for our lattice setup, which would suffer
from substantial discretization artifacts with a sharp cutoff.
We choose $R_{C} < R$ (the range of the short-range potential $V$) so that we
can interpret the short-distance Coulomb regularization as
merely a redefinition of $V$, which we otherwise choose
as attractive local Gaussian potentials, $V(r) = V_0 \ee^{{-}{r^2}/{R^2}}$.
We use $R=1$ and $R_C^2 = 0.1$ in all calculations and we emphasize that
the actual concrete form of the potential at small $r$ is irrelevant because the
expression for $\Delta E(L)$ is universal and does not depend on any particular
choice for the short-range interaction.

\begin{figure}[tbhp]
 \includegraphics[width=0.9\columnwidth]{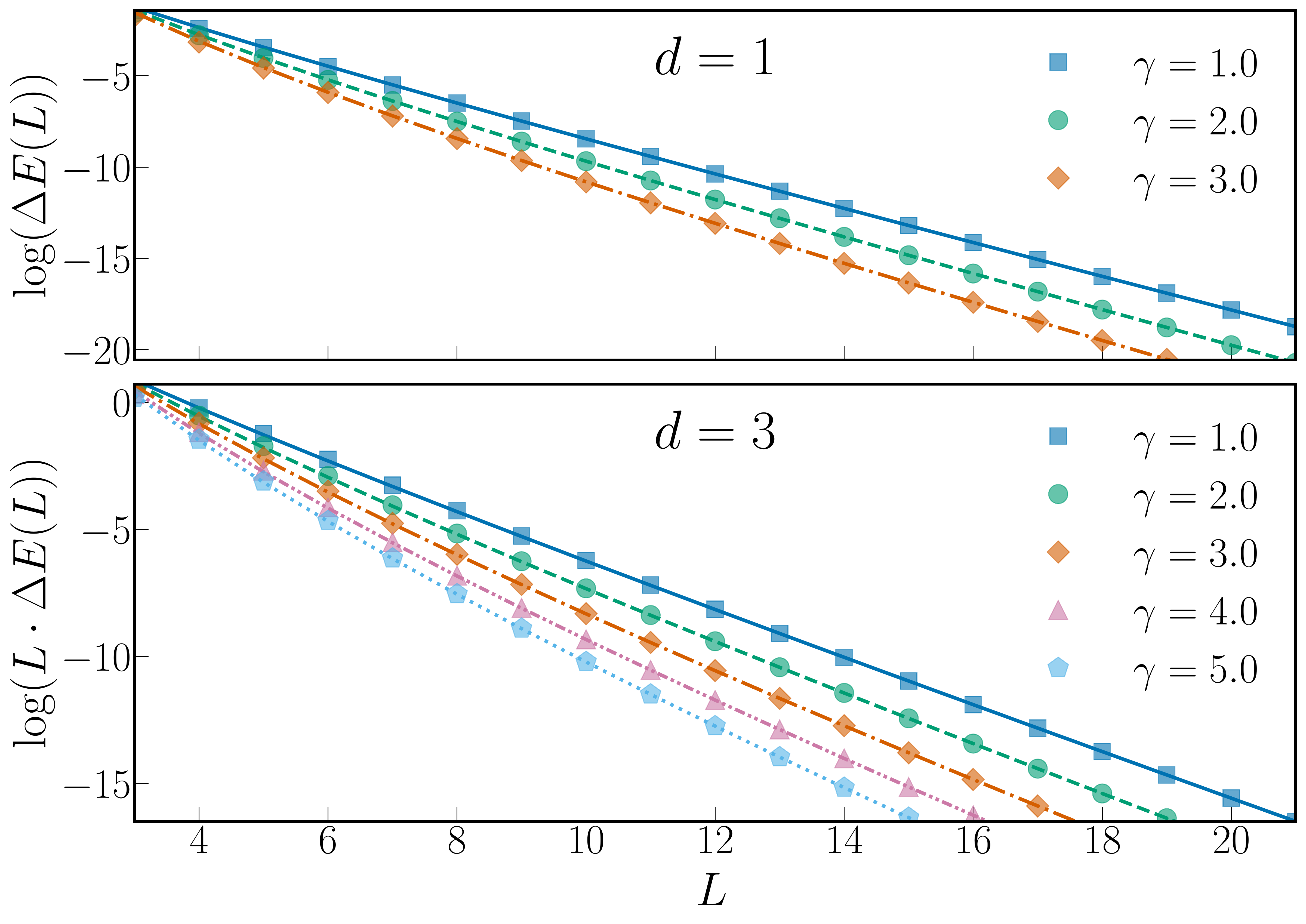}
 \caption{%
  Finite-volume energy shifts for two-body system with
  attractive Gaussian interaction plus repulsive
  Coulomb force of varying strength $\gamma$ in one (upper panel)
  and three (lower panel) dimensions.
  For each $\gamma$ the attraction is adjusted to fix the
  ground state at $\kappa_\infty \approx 0.86$ in infinite volume.
  All quantities are given in units of the mass $m=1$ (see text).
 }
 \label{fig:1d-3d}
\end{figure}

In Fig.~\ref{fig:1d-3d} we show our numerical results for 1D
and 3D systems (in the upper and lower panels, respectively).
For each case we use a logarithmic scale for the vertical axis, and we have
scaled $\Delta E(L)$ with a factor $L$ for the 3D results to account for the
overall factor $1/L$ in Eq.~\eqref{eq:DeltaE-3D-final}.
For each choice of $\gamma$, we adjust the strength of the attractive Gaussian
potential to keep the binding momentum $\kappa_\infty$ roughly constant in
infinite volume.
The precise values for $V_0$ that we use for the different cases are given
in the Supplemental Material~\cite{SuppMat}.
In the limit $\gamma\to0$, our expressions for $\Delta E(L)$ reduce to
the known results for systems with only short-range interaction, and with the
axis scalings chosen in Fig.~\ref{fig:1d-3d} the data points
would follow straight lines in that case.
We can clearly observe how the actual behavior deviates more and more from the
straight-line behavior with increasing $\gamma$, highlighting the importance
of knowing the precise analytical form of the volume dependence in the presence
of Coulomb repulsion.
Indeed, the lines in the figures show the results of fitting
Eqs.~\eqref{eq:DeltaE-1D-final} and~\eqref{eq:DeltaE-3D-final} to the numerical
data, with excellent agreement.
In Table~\ref{tab:Results} we show the values for $\kappa_\infty$ and
$A_\infty$ that we extract from the FV fits for the various cases.
The uncertainties quoted were obtained by varying the volume range for the fits
as indicated in the table, and for all cases we have approximated
$E_\infty$ from an $L = 32$ calculation.
For comparison, the table includes also binding momenta and ANCs determined
with direct continuum calculations (explained further in the Supplemental
Material~\cite{SuppMat}).
We find excellent agreement up to minor deviations, which we attribute
primarily to small discretization errors in the lattice data.
For 1D, we used a lattice spacing $a_{\text{latt}} = 1/30$, while the 3D
calculations were performed with $a_{\text{latt}} = 1/6$.
The finer lattice spacing required for good accuracy in 1D is an artifact of
using the 3D Coulomb potential in a 1D setup, which causes
significant sensitivity to the regularization of the Coulomb singularity at
$r=0$.
In 3D, the reduced radial wave function vanishes at $r=0$, which suppresses
sensitivity to regularization details.
In practical applications for example to analyzing Lattice EFT data,
one would expect other uncertainties to dominate.
Overall, our numerical calculations confirm conclusively that FV fits can be
used for accurate ANC determinations even in the presence of strong Coulomb
repulsion.

\begin{table}[htbp!]
  \def\arraystretch{1.2}
  \pgfplotstabletypesetfile[
  columns = {
   gamma,
   kappae, Ance, range,
   kappac, Ancc
  },
  columns/gamma/.style = {
   string type, column name=$\gamma$, column type={c|}
  },
  columns/kappae/.style = {
   string type, column name=$\kappa_\infty$,
   column type={>{\centering\arraybackslash}p{5.7em}}
  },
  columns/Ance/.style = {
   string type, column name=$A_\infty$,
   column type={>{\centering\arraybackslash}p{4.0em}}
  },
  columns/kappac/.style = {
   string type, column name=$\kappa_\infty$,
   column type={>{\centering\arraybackslash}p{3.6em}}
  },
  columns/Ancc/.style = {
   string type, column name=$A_\infty$,
   column type={>{\centering\arraybackslash}p{3.6em}}
  },
  columns/range/.style = {
   string type, column name={$L$ range},
   column type={>{\centering\arraybackslash}p{4.5em}|}
  },
  every head row/.style= {
   before row = {
    \TopRule
    & \multicolumn{3}{c|}{Finite-volume fit}
    & \multicolumn{2}{c}{\ Continuum result\ }
    \\
    \MidRule
   },
   after row = \MidDoubleRule
  },
  every row no 0/.style = {
   before row = {
    \multicolumn{6}{c}{$d=1$} \\ \MidRule
   },
  },
  every row no 3/.style = {
   before row = {
    \hline
    \multicolumn{6}{c}{$d=3$} \\ \MidRule
   },
  },
  every last row/.style = {after row = \BottomRule}
 ]{results.txt}
 \caption{%
  Fit results for the calculations shown in Fig.~\ref{fig:1d-3d}.
  All quantities are given in units of the mass $m=1$ (see text).
 \label{tab:Results}
 }
\end{table}
\begin{figure}[tbhp]
 \includegraphics[width=0.9\columnwidth]{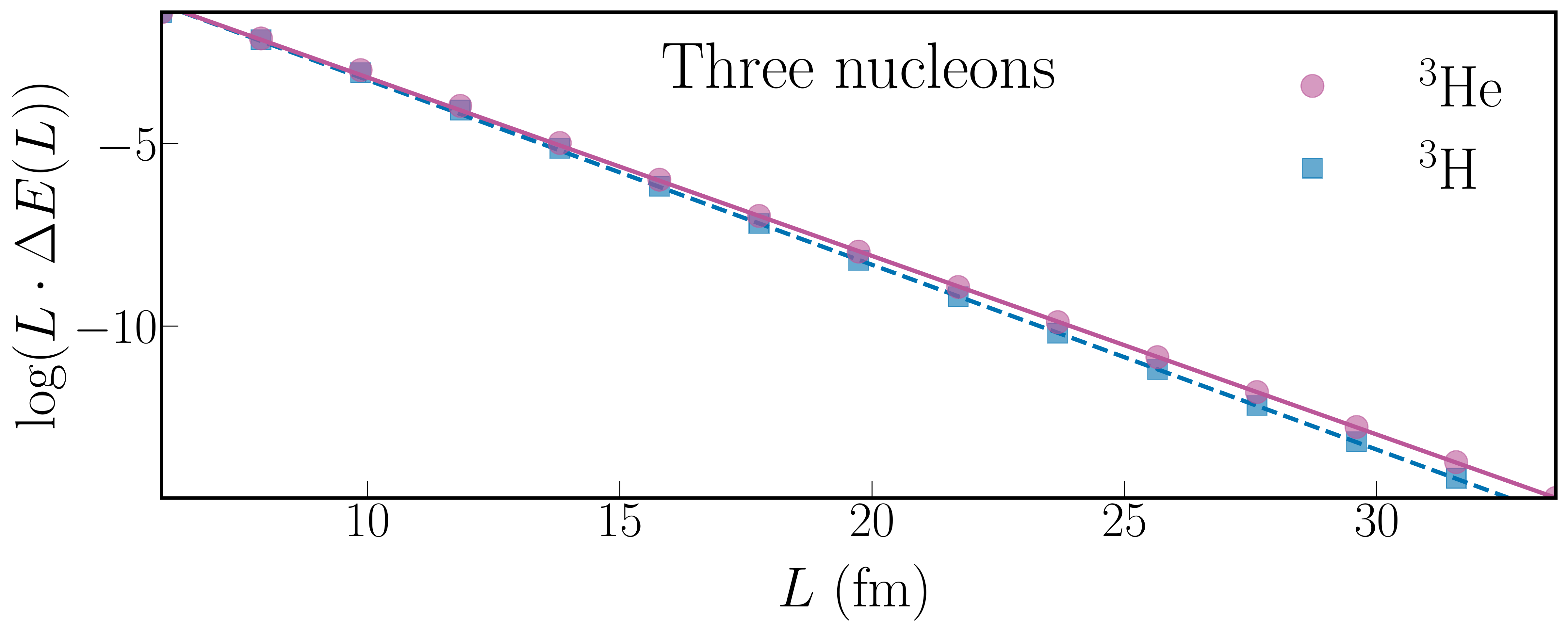}
 \caption{%
  Finite-volume energy shifts for \isotope[3]{He} and \isotope[3]{H}
  calculated with an $SU(4)$-invariant contact interaction (see text for
  details).
 }
 \label{fig:3N}
\end{figure}

We furthermore consider a three-nucleon system on a lattice with
SU(4)-invariant contact interaction that corresponds to a Pionless EFT
description (see Ref.~\cite{Hammer:2019poc} for a review of nuclear EFTs).
For this calculation, which we describe in more detail in the Supplemental
Material~\cite{SuppMat},
we tune a two-body single-lattice-site coupling to
produce a deuteron with binding energy $1$ MeV (in infinite volume), and in addition
a three-body single-lattice-site coupling to reproduce the triton at its physical
binding energy.
For \isotope[3]{He} we add a repulsive Coulomb interaction between protons and
tune a short-distance parameter associated with a $pp$ contact interaction to
reproduce the \isotope[3]{He} binding energy in infinite volume.
In Fig.~\ref{fig:3N} we show the energy shift for both the \isotope[3]{H}
and \isotope[3]{He} energies for a range of volumes between 3 and 17 lattice units,
with infinite-volume energies approximated by $L_\infty \approx 39.5$ fm
(20 lattice units).
We fit the numerical results with the analytic expression for the volume
dependence and find that already for a system with relatively weak Coulomb
repulsion like \isotope[3]{He} it is important for the $pd$ breakup channel
to make use of the relation derived in this work, which for general two-cluster
states is the leading term in multipole expansions of the permanent and/or
induced charge densities.
In particular, the ANC extracted as $A_\infty = 1.44(1)~\text{fm}^{{-}1/2}$
would be off by about 5\% if we were to fit the volume dependence assuming
neutral clusters, already for $\gamma$ as small as
$0.046~\text{fm}^{{-}1}$ here.
This importance will increase for systems with more protons and/or weaker binding
energy, such as proton halo nuclei.
As an additional check, we can compare the ANC that we obtain from fitting the
energy volume dependence to a direct extraction from the finite-volume wave
function (as discussed in Ref.~\cite{Konig:2017krd} and described further in the
Supplemental Material~\cite{SuppMat}).
From that procedure we obtain a value of $1.46(1)~\text{fm}^{{-}1/2}$, in good
agreement with the result from the energy fit.
Moreover, taking into account the SU(4)-symmetric scenario that we consider
here, our ANC is in reasonably good agreement with Ref.~\cite{Friar:1988zzb},
which reports an $S$-wave ANC for \isotope[3]{He} of
$1.82(1)~\text{fm}^{{-}1/2}$, converted to our conventions.

\paragraph{Summary and outlook}
We have derived the leading finite-volume energy correction for bound states of
two charged clusters in a periodic volume.
Numerical calculations for several examples in 1D and 3D systems give confidence
that the analytical results are correct.
Our derivation does not rely on perturbation theory and is therefore applicable
to medium-mass and heavy nuclei with large numbers of protons.
The results should be immediately useful for \textit{ab initio} calculations of
bound nuclei in periodic boxes, which is standard practice in Lattice EFT
calculations.
The form of the finite-volume energy correction provides a better extrapolation
for the binding energy at infinite volume.
Even more importantly, it yields a new efficient method for calculating
ANCs that play an important role in low-energy astrophysical capture reactions.
The extension to bound states of charged particles with nonzero angular momentum
is currently under investigation.

\begin{acknowledgments}
The work of H.Y.\ and S.K.\ was supported in part by the National Science
Foundation under Grant No. PHY--2044632.
D.L.\ is supported by the U.S.\ Department of Energy (No.\ DE-SC0013365
and No.\ DE-SC0021152) and the Nuclear Computational Low-Energy Initiative
(NUCLEI) SciDAC-4 project (DE-SC0018083).
This material is based upon work supported by the U.S. Department of Energy,
Office of Science, Office of Nuclear Physics, under the FRIB Theory Alliance,
Award No.\ DE-SC0013617.
\end{acknowledgments}

\appendix

\section*{Supplemental Material}

\newcommand{\smsection}{\subsection*}
\newcommand{\smsubsection}{\subsubsection*}

\begin{widetext}

\smsection{Full derivation of the energy shift in one dimension}
\label{sec:Direct-1D}

In this section we provide a detailed derivation of the energy shift in one
dimension quoted in the main text.
Despite studying a two-particle bound state restricted to a 1D finite ``volume''
(\ie, a circle), this setup keeps the Coulomb potential to have its normal
three-dimensional form.
The purpose of this choice is to illustrate the complications arising from the
long-range interaction in a simplified setup.
The equation describing the bound state of interest is equivalent to the radial
Schrödinger equation for a 3D $S$-wave state, only that the boundary condition
at the origin is different.
Importantly, this means that standard results from $S$-wave scattering theory for
charged particles apply, and consequently we can make use of these in the
derivation.

\smsubsection{No Coulomb potential}

To explain the method, it is instructive to first consider the case without
Coulomb interaction.
The bound-state wave function for the state defined on a circle with
circumference $L$ then takes the general form
\begin{equation}
 u(r) = A(\ii\kappa) \exp({-}\kappa r) + B(\ii\kappa) \exp({+}\kappa r) \,,
 \label{eq:psi-AB}
\end{equation}
where $\kappa = \kappa(L)$.
In infinite volume we have $\kappa_\infty = \kappa(L=\infty)$ and
\begin{subalign}
 A(\ii\kappa_\infty) &= A_\infty \,, \\
 B(\ii\kappa_\infty) &= 0 \,,
\end{subalign}
since the asymptotic wave function must have a pure exponential fall-off with ANC
$A_\infty$.
The reason for including a factor $\ii$ in the argument of $A$ and $B$ will
become apparent below.

Limiting our discussion for the time being to states with even parity, the
derivative of $\psi(r)$ needs to vanish at the box boundary in order to satisfy
the periodic boundary condition.
Setting $\psi'(r=L/2) = 0$ and Taylor-expanding this condition around
$\kappa = \kappa_\infty$ yields
\begin{equation}
 \Delta \kappa(L) = \frac{A(\ii\kappa_\infty)}{B'(\ii\kappa_\infty)}
 \exp({-}\kappa L) + \mathcal{O}\big(\exp({-}2\kappa L)\big) \,,
 \label{eq:Delta-kappa-bare}
\end{equation}
where $B'(\ii\kappa) = {\dd B(\ii\kappa)}{\dd\kappa}$.
The prefactor in Eq.~\eqref{eq:Delta-kappa-bare} can be determined from the
analytic structure of the wave function as a function of general complex momentum
$k$.
Following Ref.~\cite{Faldt:1996sd} (with slightly adjusted notation), we can
write $\psi(r) = w(\ii\kappa,r)$ with
\begin{equation}
 w(k,r) = \sqrt{\kappa + \ii k} \left[
  \ee^{\ii \delta(k)} f(k,r) - \ee^{-\ii \delta(k)} f(-k,r)
 \right] \,,
\end{equation}
where $f(\pm k,r) \sim \exp(\pm kr)$ for $r\to\infty$ are the Jost solutions.
Comparing with Eq.~\eqref{eq:psi-AB}, we read off
\begin{align}
 A(k) &= \sqrt{\kappa + \ii k} \, \ee^{\ii \delta(k)} \,, \\
 B(k) &= {-}\sqrt{\kappa + \ii k} \, \ee^{{-}\ii \delta(k)} \,,
\end{align}
and as expected we have that~\cite{Faldt:1996sd}
$A(k \to \ii\kappa) = A_\infty$ and $B(k \to \ii\kappa) = 0$.
To calculate the derivative of $B$ at the pole, we use the parametrization of
the S-matrix $s(k)$ given in Ref.~\cite{Faldt:1996sd}:
\begin{equation}
 s(k) = \ee^{2\ii\delta(k)}
 = \frac{\left[A_\infty G(k)\right]^2}{\kappa + \ii k} \,,
\end{equation}
with $G(\ii\kappa) = 1$.
This yields $B(k) = {-}(\kappa + \ii k)/(A_\infty G(k))$, and then
\begin{equation}
 \frac{\dd B(k)}{\dd k}\Bigg|_{k=\ii\kappa} = {-}\ii/A_\infty \,.
\end{equation}
With $\dd/\dd k = -\ii\times\dd/\dd\kappa$ it follows that $B'(\ii\kappa_\infty)
= 1/A_\infty$, so that overall we arrive at
\begin{equation}
 \frac{A(\ii\kappa_\infty)}{B'(\ii\kappa_\infty)} = A_\infty^2 \,.
\end{equation}

\smsubsection{Repulsive Coulomb force}
\label{sec:DirectCoulomb}

We now assume that a repulsive Coulomb force as defined in main text
is present in addition to the short-range interaction $V$.
For this case we can write the general wave function as
\begin{equation}
 w(\eta,r) = \sqrt{\kappa + \ii k} \left[
  \ee^{\ii\delta(k)} f^{(+)}(\eta,r)
  - \ee^{-\ii\delta(k)} f^{(-)}(\eta,r)
 \right] \,,
 \label{eq:w-Coulomb}
\end{equation}
with $\eta = \gamma/(2k)$, and using Coulomb Jost solutions $f^{(\pm)}(\eta,r)$
that satisfy~\cite{Vaandrager:2019lje}
\begin{subalign}[eq:Coulomb-Jost]
 f^{(+)}(\eta,r) &\sim \ee^{\pi\eta/2} W_{{-}\ii\eta,\frac12}({-}2\ii kr) \,, \\
 f^{(-)}(\eta,r) &\sim \ee^{\pi\eta/2} W_{\ii\eta,\frac12}(2\ii kr)
\end{subalign}
for $r\to\infty$.
Making an ansatz for $\psi(r)$ analogous to Eq.~\eqref{eq:psi-AB}, but with
$f^{(\pm)}(\eta,r)$ replacing the simple exponential functions, now yields for
the finite-volume shift of $\kappa$:
\begin{equation}
 \Delta \kappa(L)
 = \frac{A(\ii\kappa_\infty)}{B'(\ii\kappa_\infty)}
 \frac{\frac{\dd}{\dd r}f^{(+)}({-}\ii\bar{\eta},r)}
 {\frac{\dd}{\dd r}f^{(-)}({-}\ii\bar{\eta},r)}\Bigg|_{r=L/2}
 + \cdots \,.
 \label{eq:Delta-kappa-bare-C}
\end{equation}
Note that in order to keep the notation simple we use here the same symbols
as in the previous section and do not add annotations that indicate they are
modified by the presence of the Coulomb force.
Higher-order contributions to $\Delta\kappa$ are indicated with ``$\cdots$'' in
Eq.~\eqref{eq:Delta-kappa-bare-C}; their precise form will be discussed below.

We proceed now as before.
The phase shift $\delta(k)$ in Eq.~\eqref{eq:w-Coulomb} collects the
effects from the total (Coulomb plus short-range) interaction.
It is customary to separate out the pure Coulomb phase shift $\delta_c(k)$,
defined most carefully~\cite{Dzieciol:1999aa} via
\begin{equation}
 \ee^{\ii\delta_c(k)}
 = \left(\frac{\Gamma(1+\ii\eta)}{\Gamma(1-\ii\eta)}\right)^{\!1/2} \,,
\end{equation}
and write
$\delta(k) = \delta_c(k) + \tilde{\delta}(k)$.
Using these definitions we can state that
near the bound-state pole the S-matrix behaves like~\cite{Blokhintsev:2016mie}
\begin{equation}
 \ee^{2\ii\delta(k)}
 \sim \frac{A_\infty^2 \ee^{\ii\pi\bar{\eta}}}{\kappa + \ii k}
 + \text{regular terms} \,,
\end{equation}
where we have introduced $\bar{\eta} = \gamma/(2\kappa)$, \ie, $\eta(k) =
{-}\ii\bar{\eta}$ for $k=\ii\kappa$.
(Alternatively we can proceed as above and collect the regular terms in a
function $G(k)$ that satisfies $(G(\ii\kappa) = 1$.)

Compared to the expression without Coulomb interaction there is a peculiar
additional phase factor in the residue.
It turns out that this phase is in fact crucial to obtain the expected behavior
for the bound-state wave function.
To see this, we read off
\begin{equation}
 A(k) \sim \sqrt{\kappa + \ii k} \, \ee^{\ii\delta(k)}
 \to A_\infty \ee^{\ii\pi\bar{\eta}/2}
\end{equation}
for $k\to\ii\kappa$, and find that the extra phase cancels precisely against
the phase factor in the Jost solution $f^{(+)}(\eta,r)$.
Overall, since exactly as in the case without Coulomb interaction we immediately
see
\begin{spliteq}
 B(k) &\sim (\kappa + \ii k) \ee^{{-}\ii\delta(k)} \\
      &= (\kappa + \ii k) \frac{\ee^{{-}\ii\pi\bar{\eta}/2}}{A_\infty}
 \to 0 \mathtext{for} k\to\ii\kappa \,,
\end{spliteq}
this leads to the expected asymptotic form for the bound-state wave function:
\begin{equation}
 w({-}\ii\bar{\eta},r) \sim A_\infty W_{{-}\bar\eta,\frac12}(2\kappa r) \,.
\end{equation}

We can also calculate the derivative of $B(k)$ and find
\begin{equation}
 \frac{A(\ii\kappa_\infty)}{B'(\ii\kappa_\infty)}
 = A_\infty^2 \ee^{\ii\pi\bar{\eta}} \,.
\end{equation}
For the energy shift this gives
\begin{equation}
 \Delta\kappa(L)
 = A_\infty^2 \ee^{\ii\pi\bar{\eta}}
 \frac{\frac{\dd}{\dd r}f^{(+)}({-}\ii\bar{\eta},r)}
 {\frac{\dd}{\dd r}f^{(-)}({-}\ii\bar{\eta},r)}\Bigg|_{r=L/2}
 + \cdots \,.
\end{equation}
Finally, we can also insert Eqs.~\eqref{eq:Coulomb-Jost} for the Jost solution
and notice that the terms in the ellipsis are bounded by the decaying solution
evaluated at larger distances.
Overall, we arrive at
\begin{equation}
 \Delta\kappa(L)
 = A_\infty^2 \ee^{\ii\pi\bar{\eta}}
 \frac{W'_{{-}\bar{\eta},\frac12}(\kappa L)}
 {W'_{\bar{\eta},\frac12}({-}\kappa L)}
 + \OO\left[\ee^{-2\kappa L}\right] \,,
 \label{eq:Delta-kappa-final}
\end{equation}
where the prime denotes the derivative of the Whittaker function with respect
to its argument.
Note that correction terms are in fact also Whittaker functions (and/or
derivatives thereof), but for simplicity we write them as
$\OO\left[\ee^{-2\kappa L}\right]$; since they actually fall of faster than this
exponential, this is permissible.
Again we point out that phase in the prefactor in
Eq.~\eqref{eq:Delta-kappa-final} is indeed
crucial: while $W'_{{-}\bar{\eta},\frac12}(\kappa L)$ is real,
$W'_{\bar{\eta},\frac12}({-}\kappa L)$ is not, but its phase is exactly given by
$\ee^{\ii\pi\bar{\eta}}$~\cite{Slater:1960} and thus cancels out,
leaving $\Delta\kappa(L)$ real overall.
The result for $\Delta E(L)$ quoted in the main text is obtained by noting that
$\Delta E(L) = \kappa_\infty \Delta\kappa(L) / \mu$.

\smsection{Full derivation of the energy shift in three dimensions}

Here we present the full derivation of the finite-volume energy shift in
three dimensions, starting from following expression in the main text:
\begin{equation}
 \EdL(L) = \sum_{|\vec n|=1} \int_{B^3} \dd^3 x \,
 \psip_L(\vec x) V(\vec x)\psip_L(\vecx-\vecn L)
 + \OO(\ee^{{-}\sqrt{2}\kappa L}) \,.
\label{eq:EdL-1}
\end{equation}
We demonstrate how from this we obtain the final form for $\Delta E(L)$ shown
in the main text by considering explicitly the term with $\vec n = {-}\hat z$
and exploiting the symmetry of the wave function.
Introducing a factor $C_3=6$ to account for all terms in Eq.~\eqref{eq:EdL-1},
we can write
\begin{equation}
 \EdL(L) = 6 \int_{B^3} \dd^3 x \,
 \psip_L(\vec x) V(\vec x)\psip_L(\vecx-\hat{z}L)
 + \OO(\ee^{-\sqrt{2}\kappa L}) \,.
\label{eq:EdL-2}
\end{equation}
For $\vec{x}\in B^3$, we have:
\begin{equation}
 [H_0 + V_{C,\{L\}}(\vecx) + V(\vecx)] \psip_L(\vec x-\hat z L)
 = {-}\tilde{E}(L) \psip_L(\vec x-\hat z L) +
 V(\vec x )\psip_L(\vec x-\hat z L) \,.
\end{equation}
This is similar to the derivation for systems with short-range interactions
only~\cite{Luscher:1985dn}, but it differs in the details due to our
perturbative treatment of $\Delta V_C$: instead of $\ket{\psi_\infty}$ as
eigenstate of $H$, we are dealing here with $\ket{\tilde{\psi}_L}$ as eigenstate
of $\tilde{H}_L$.
We integrate by parts along each coordinate axis, using
\begin{equation}
 v \nabla^2 u
 = u \nabla^2 v + \nabla \cdot (v \nabla u - u \nabla v)\,,
\label{eq:simp-int3d}
\end{equation}
and after that we now use
$[H_0 + V_{C,\{L\}}(\vecx) + V(\vecx)+\tilde{E}(L) ]\psip_L(\vecx)=0$ to
eliminate the term $u \nabla^2 v$.
The second term in Eq.~\eqref{eq:simp-int3d} is a divergence and
we can therefore use Gauss's theorem to obtain
\begin{multline}
 \EdL(L)
 = 6\int_{B^3} \dd^3x \, \psip_L(\vecx)
  \big[H_0 + V_{C,\{L\}}(\vecx) + V(\vecx)+\tilde E(L)\big]
  \psip_L(\vecx-\hat{z}L) \\
 = \frac{6}{\mu} \sum_{(ijk)} \int_{B^2_{\pm k}} \dd x_i\dd x_j \,
  \psip_L(\vecx-\hat z L)
  \frac{\partial}{\partial x_k}\psip_L(\vecx)
 + \OO(\ee^{{-}\sqrt{2}\kappa L}) \,,
\label{eq:int3d-perm}
\end{multline}
where the $\sum_{(ijk)}$ now indicates a sum over all permutations
of the coordinate indices $(x_i,x_j,x_k)=(x,y,z)$, and
\begin{equation}
 B^2_{\pm k}
 = \{\vec{x}\in\mathbb{R}^3|-L/2\leq x_i,x_j\leq L/2,\ x_k = \pm L/2\}
\label{eq:I2}
\end{equation}
denotes the square face at $x_k=\pm L/2$ of the cube centered at the
origin.
From the asymptotic behavior of the infinite-volume wave function $\psiinf$,
which carries over to $\psip_L(\vecx)$ up to negligible corrections, it is
clear that terms in Eq.~\eqref{eq:int3d-perm} with $k\neq z$ are of higher
exponential order.
Therefore, writing $B^2_{\pm k} \equiv B^2_{\pm x_k}$ to slightly simplify the
notation in the following, we get
\begin{spliteq}
 \EdL(L) = \frac{6}{\mu}\int_{B^2_{+z}}
 \dd x\dd y \, \psip_L(\vecx-\hat{z}L)
 \frac{\partial}{\partial z}\psip_L(\vecx)
 + \OO(\ee^{{-}\sqrt{2}\kappa L}) \,.
\label{eq:intparts3d}
\end{spliteq}

\begin{figure}[hbt!] \centering
 \includegraphics[width=0.4\textwidth]{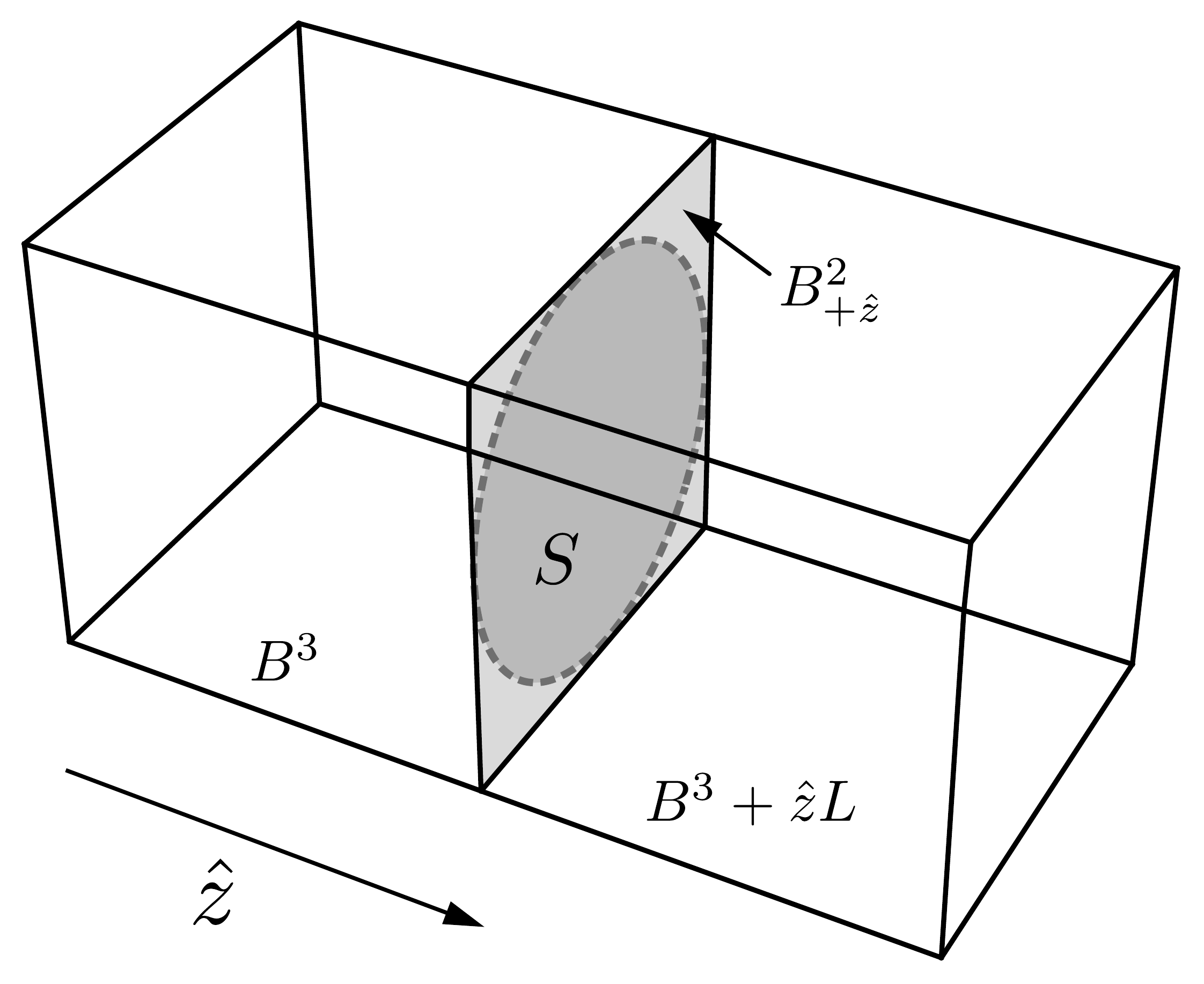}
 \caption{Visualization of various integration regions.
  In this picture we display two periodic cubes with $O=(0,0,0)$ centered
  in left cube $B^3=[-L/2,L/2]^3$. $B^2_{+z}$ is the gray square surface
  region and $S$ is the deeper gray surface area.
  The surface element $\dd S$
  is the small area to be integrated around $\theta=0$ in spherical
  coordinates, and one can use either the first fundamental form or simply the
 area of $S$ to compute $\dd S$ (see text).}
 \label{fig:CubicRegions}
\end{figure}

We now change coordinates as illustrated in Fig.~\ref{fig:CubicRegions} to
compute the surface integral
\begin{equation}
 \int_{B^2_{+z}} \dd x \dd y = \int_{B^2_{+z}} \dd\phi\dd\theta \sqrt{g} \,,
\end{equation}
since for a disk $S$ located in the $z=L/2$ plane the surface element
can be written as
\begin{equation}
 \dd S = \frac{\pi}{4} \dd\big(L^2\tan^2\theta \big) \,.
\end{equation}
We have $\sqrt{g} = r^2\tan{\theta}$ via the first fundamental form, and
$r\cos{\theta}=L/2$ at the edge.
We can also derive this $\dd S$ from the surface area of the disk $S$,
\begin{equation}
 S = \pi\left(\frac L2\tan\theta\right)^2 \,.
\end{equation}
To leading order in the perturbative expansion
we have $\psip_L(\vecr_0) = \psiinf(\vecr_0)$, with $S$-wave symmetry, and with
\begin{equation}
 \frac{\partial}{\partial z}=\cos\theta \frac{\partial}{\partial r}
\end{equation}
we then get
\begin{spliteq}
 \EdL(L)
 &= \frac{6}{\mu}\int_{B^2_{+z}} \dd \phi \dd \theta \sqrt{g} \cos\theta
  \,\psip_L(\vecr-\hat zL)\frac{\partial}{\partial z}
  \psip_L(\vecr)
  + \OO(\ee^{{-}\sqrt{2}\kappa L}) \\
 &= \frac{12\pi}{\mu} \int_{L/2}^{L/\sqrt2} \dd\!\left(\frac{L}{2 r}\right)
  r^2\psiinf(r)\frac{\dd}{\dd r}\psiinf(r)
  + \Edp'(L)
  + \OO(\ee^{{-}\sqrt{2}\kappa L}) \,,
\label{eq:withsphericalsym}
\end{spliteq}
We have introduced here the additional correction term $\Edp'(L)$ to
account for the higher-order difference between $\ket{\psip_L}$ and
$\ket{\psiinf}$.
The integrand in the remaining area
$B^2_{+z}-S$, shown in Fig.~\ref{fig:CubicRegions},
only contributes to higher-order terms $\OO(\ee^{{-}\sqrt{2}\kappa L})$,
which follows again from the asymptotic behavior of the wave function.
This contribution can therefore be neglected and the right-hand side can now be
explicitly integrated, yielding
\begin{equation}
 {-}\frac{2\pi L}{\mu} \int_{L/2}^{L/\sqrt2}
 \dd r \, \psiinf(r)\frac{\dd}{\dd r}\psiinf(r)
 = {-}\frac{\pi L}{\mu}\psiinf(L/2)^2
 + \OO(\ee^{{-}\sqrt{2}\kappa L}) \,,
\label{eq:withsphericalsym-2}
\end{equation}
and overall we arrive at
\begin{equation}
 \Delta E(L) = {-}\frac{3\pi L}{\mu}{\psiinf}(\vec r_0)^2 + \Edp(L)
 + \Edp'(L)+\OO(\ee^{{-}\sqrt{2}\kappa L}) \,.
 \label{eq:DeltaE-3D-final2}
\end{equation}
Finally, we can insert the known asymptotic form of $\psiinf(r)$
and obtain
\begin{equation}
 \Delta E(L) = \underbrace{%
  {-}\frac{3A_\infty^2}{\mu L}
  \Big[W_{-\bar\eta,\frac{1}{2}}(\kappa L)\Big]^2
 }_{\equiv \Delta E_0(L)}
 + \Edp(L)
 + \Edp'(L)
 + \OO(\ee^{{-}\sqrt{2}\kappa L})
\end{equation}
for the overall energy shift at volume $L$ in three dimensions, as stated in
the main text.

\medskip
The term labeled $\Delta E_0(L)$ above could be obtained from a direct
generalization of the overlap method presented in Ref.~\cite{Luscher:1985dn}.
In this calculation one would replace $\psip_L$ with $\psiinf$ in
Eq.~\eqref{eq:int3d-perm}.
This amounts to ignoring all higher-order corrections in the perturbative
expansion and therefore naturally recovers the same leading term.
The key advantage of our approach is that we obtain explicit expressions for the
otherwise unknown subleading corrections, which we proceed to analyze in the
following.

\smsubsection{First correction term}

To understand the first correction term $\Edp(L)$, defined in the main text
and included in Eq.~\eqref{eq:DeltaE-3D-final2} above, we need to consider the
first-order correction to the energy eigenvalue, given by
\begin{equation}
 \Edp^{(1)}(L) = \mbraket{\psiinf}{\Vd_C}{\psiinf}
 \sim\frac{|\gamma_{\infty}|^2}{4\pi}\int_{\RR^3} \dd^3x
 \frac{1} {|\vec x|^2} \big[W_{{-}\bar\eta,\frac{1}{2}}
 \big(2\kappa (|\vec x|)\big)\big]^2 \Vd_{C}( \vec x) \,.
\label{eq:correctionterma}
\end{equation}
This expression can be viewed as variational correction by treating
$\psiinf$ as trial functions and we therefore have that
\begin{equation}
\Edp^{(1)}(L) - E_\infty=   \mbraket{\psiinf}{H+\Vd_C}{\psiinf}
=\mbraket{\psiinf}{\tilde H_L}{\psiinf} \geq -\tilde E(L) \,.
\end{equation}
With $\Vd_C$ being semi-positive definite, we also have
that $ \mbraket{\psip_L}{\Vd_C}{\psip_L}\geq 0$, which consequently implies
\begin{equation}
 {-}\tilde E(L)
 = \mbraket{\psip_L}{H+\Vd_C}{\psip_L}
 \geq \mbraket{\psip_L}{H}{\psip_L}\geq {-}E_\infty \,.
\end{equation}
Overall we can bound the correction $\Edp(L)$ as
\begin{equation}
 0\leq\Edp(L) \leq \Edp^{(1)}(L) \,.
\label{eq:strictbound3d}
\end{equation}

We proceed to analyze the asymptotic behavior of $\Edp^{(1)}(L)$ because as
we established above, this governs the overall scaling of $\Edp(L)$.
It is simple to check that
\begin{spliteq}
 \Edp^{(1)}(L)
 &= \frac{|\gamma_{\infty}|^2}{4\pi}\int_{\RR^3} \dd^3x
 \frac{1} {|\vec x|^2} \big[W_{{-}\bar\eta,\frac{1}{2}}
 \big(2\kappa (|\vec x|)\big)\big]^2 \Vd_{C}( \vec x) \\
 &=\frac{|\gamma_{\infty}|^2}{4\pi}\int_{\RR^3-B^3} \dd^3x
 \frac{1} {|\vec x|^2} \big[W_{{-}\bar\eta,\frac{1}{2}}
 \big(2\kappa (|\vec x|)\big)\big]^2 \Vd_{C}( \vec x)\\
 &=\frac{3|A_\infty|^2}{2\pi}\int_{U_L} \dd^3 x \frac{1}{|\vecx|^2}
 \big[W_{{-}\bar\eta,\frac{1}{2}}\big(2\kappa(|\vecx|)\big)\big]^2
 \Vd_{C,(0,\hat z)L}(\vecx) + \OO(\ee^{{-}\sqrt2\kappa L}) \,.
 \label{eq:1storder_ptE}
\end{spliteq}
The second line uses the support of $\Vd_C(\vecx)$ is in fact $\RR^3-B^3$, and
in the last line $\OO(\ee^{-\sqrt{2}\kappa L})$ correction absorbs contribution
in region $|\vecx|\geq\sqrt2L/2$,
and $U_L$ is therefore defined as
\begin{equation}
 U_L = (B^3+\hat z L) - \{\vec x' : \abs{\vec x'} \geq L/\sqrt{2} \} \,,
\end{equation}
where $B^3+\hat z L$ denotes the volume $B^3$ shifted by $L$ in the positive
$z$ direction.
This truncation of integral volume avoids confusion caused by
singularities of $\Vd_C$, which now lie outside of $U_L$.

The key step is now to notice that the integrand in Eq.~\eqref{eq:1storder_ptE}
remains non-negative on $U_L$, and that therefore it is possible to find
functions $g_-$ and $g_+$ such that
\begin{equation}
 \int_{U_L} \dd^3\,\vec x\,g_-(\vec x)
 \leq \Edp^{(1)}(L)
 \leq\int_{U_L} \dd^3\,\vec x\,g_+(\vec x) \,.
\end{equation}
For the Whittaker functions we have that
\begin{equation}
 W_{-\bar\eta,1/2}(2\kappa r) \sim \frac{\ee^{{-}\kappa r}}
 {(\kappa r)^{\bar\eta}}
 \mathtext{for} r\to \infty \,.
\label{eq:asymptoticW}
\end{equation}
This means for large $L$ and
$\kappa r=\kappa |\vec x|>\kappa L/2$, there exists a pair of constants $C$, $C'$
such that the following inequality holds for all $\abs{\vec x} > L/2$:
\begin{equation}
 C'\frac{\ee^{-\kappa r}}{(\kappa r)^{\bar\eta}}
 \leq W_{-\bar\eta,1/2}(2\kappa r)\leq C \frac{\ee^{{-}\kappa r}}
 {(\kappa r)^{\bar\eta}} \,.
\label{eq:C-W}
\end{equation}

We proceed and derive some inequalities that will help us bound the integral in
Eq.~\eqref{eq:1storder_ptE}.
If we define $\vec r_0 = (0,0,L/2)$ and introduce a shifted coordinate vector
$\vecr$. On the integration region $U_{L}$ we have
\begin{equation}
 L/2+z+\rho^2/2L\leq|\vec{r}+\vec r_0|\leq L/2+r\,.
 \label{eq:ineq_r_on_Ul}
\end{equation}
For the left-hand side of Eq.~\eqref{eq:ineq_r_on_Ul} we have expanded
$|\vec{r}+\vec r_0|$ in cylindrical coordinate $(z,\rho,\phi)$,
\begin{equation}
 |\vec{r}+\vec r_0|= L/2+z+\rho^2/L -2 z \rho^2/L^2 +\OO(1/L^3)
 \geq L/2+z+ \rho^2/2L \,,
 \label{eq:radialineq}
\end{equation}
while the right-hand side of Eq.~\eqref{eq:ineq_r_on_Ul} is just the triangle
inequality.
In particular, we have $L/2+z\leq L/\sqrt{2}$ on $U_{L}$, and thus
\begin{equation}
 C' \bar\eta \, \frac{r \cos\theta}{L^2}
 \leq \Vd_C(r)
 \leq C \bar\eta \, \frac{z}{L^2} \,,
\label{eq:deltaVineq}
\end{equation}
with constants $C$, $C'$ independent of $L$.

With this, we can now find an upper bound for $\Edp^{(1)}(L)$.
We make use of the left-hand side of Eq.~\eqref{eq:radialineq} to bound the
exponential factor, and we furthermore bound $\Vd_C$ based on right-hand side
of Eq.~\eqref{eq:deltaVineq}:
\begin{spliteq}
 \Edp^{(1)}(L)
 & \leq \OO(\frac{\bar\eta \ee^{{-}\kappa L}}
 {(\kappa L)^{2\bar\eta+4}}) \times \int \dd z z \ee^{{-}2\kappa z}
 \times \int_0^{L/2} \dd \rho \rho \ee^{{-}\frac{\kappa\rho^2}{L}} \\
 & = \frac{\bar\eta\ee^{{-}\kappa L}}{(\kappa L)^{2\bar\eta+3}}
 \left[\OO(1) + \OO(\ee^{{-}\kappa L /2})\right] \\
 &= \OO\!\left(\frac{\bar\eta}{(\kappa L)^2}\right) \times \Delta E_{0}(L) \,.
\label{eq:DeltaEp-1-bound}
\end{spliteq}
The second step in Eq.~\eqref{eq:DeltaEp-1-bound} uses
\begin{equation}
 \int_0^{a} \dd x \, x^n \ee^{{-}\kappa b x}
 = \Gamma(n+1)/(\kappa b)^{n+1}
 + \OO(\ee^{{-}\kappa b a }) \,,
\label{eq:exp_int}
\end{equation}
and in the final step we have used
Eq.~\eqref{eq:asymptoticW}, which implies that asymptotically
$\Delta E_{0}(L) \sim \dfrac{\ee^{{-}\kappa L}}{(\kappa L)^{\bar\eta+1}}$.
Overall we have now established that
\begin{equation}
 \Edp(L) =
 \OO\!\left(\frac{\bar\eta}{(\kappa L)^2}\right) \times \Delta E_{0}(L) \,.
\label{eq:Edp-L-final}
\end{equation}

We can check this result with an explicit numerical evaluation of $\Edp(L)$.
In Fig.~\ref{fig:plotdelta} we show this for the particular (but arbitrary)
choice of parameters.
This calculation was performed using numerical integration to evaluate the
expression in Eq.~\eqref{eq:1storder_ptE} in order to approximate $\Edp(L)$,
and by using the known analytical form for the leading volume dependence as
given in Eq.~\eqref{}.
With the guiding lines that indicate inverse square and inverse cubic functions
in Fig.~\ref{fig:plotdelta} we can see that asymptotically the explicit
evaluation of the ratio $\Edp(L)/\Delta E_{0}(L)$
is nicely consistent with Eq.~\eqref{eq:Edp-L-final}.

\begin{figure}[tbhp]
 \centering
 \includegraphics[width=0.6\textwidth]{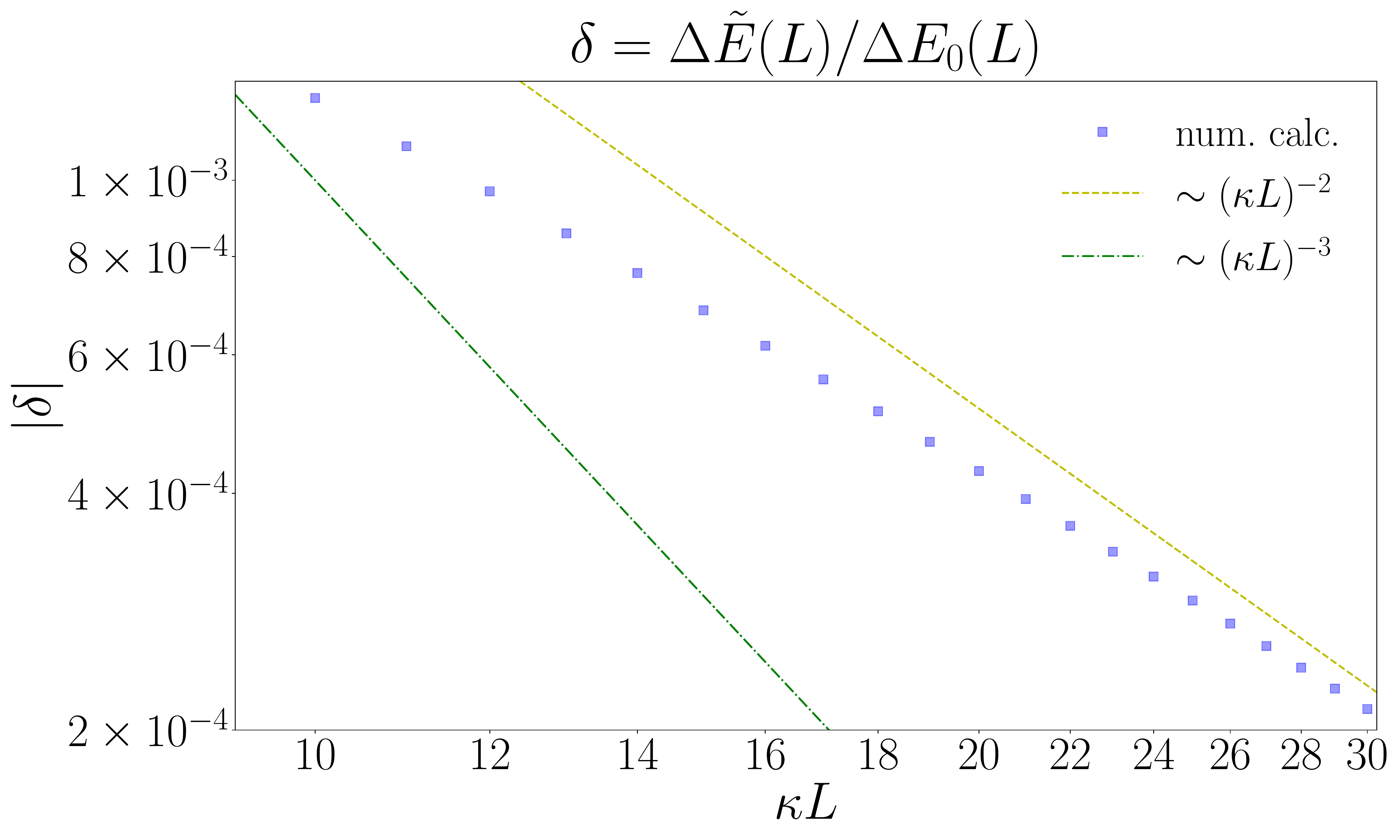}
 \caption{%
  Relative size of $\Edp(L)$ compared to $\Delta E_{0}(L)$.
  This numerical calculation, plotted as squares, was performed as described in
  the text with parameters $\gamma = 1.0, \kappa = 1.0, A_\infty = 1.0$.
  The dashed and dash-dotted line show functions $C_2 \times 1/(\kappa L)^2$
  and $C_3 \times 1/(\kappa L)^3$, respectively, with $C_2 = 0.205$ and $C_3 =
  1.0$.
  \label{fig:plotdelta}
 }
 \end{figure}

\smsubsection{Second correction term}

The second correction term in Eq.~\eqref{eq:DeltaE-3D-final2}, $\Edp'(L)$,
arises from the difference between $\ket{\psip_L}$ and $\ket{\psiinf}$.
From Eq.~\eqref{eq:DeltaE-3D-final2}, we can write out $\Edp'(L)$ explicitly as
\begin{equation}
 \Edp'(L) =\frac{6}{\mu} \int_{B^2_{+z}}\dd x \dd y
 \frac{\partial}{\partial z}\left(
  {-}2\braket{\psiinf}{\vec r}
  \left(\braket{\vec r}{\psip_L} - \braket{\vec r}\psiinf\right)
  + \left(\braket{\vec r}{\psip_L} - \braket{\vec r}\psiinf\right)^2
 \right)
 \equiv \Edp'_1(L) + \Edp'_2(L) \,.
\label{eq:DeltaE-tilde-p-explicit}
\end{equation}
We have introduced here names for the two individual contributions $\Edp'_1(L)$
and $\Edp'_2(L)$.
Their form is slightly different, but for each term it is clear that in order
to analyze its asymptotic scaling we should consider
$\ket{\psip_L} - \ket{\psiinf}$. According to
perturbation theory, we need to refer to the entire spectrum of $H$.
To this end, we define states $\ket{\psi_{\infty,klm}}$ that satisfy
\begin{equation}
 (H_0+V_C+V)\ket{\psi_{\infty,klm}} = E_k \ket{\psi_{\infty,klm}} \,,
\end{equation}
where $k$ denotes the momentum and $l,m$ are angular quantum numbers.
It suffices to study the difference $\ket{\psip_L} - \ket{\psiinf}$ in
first-order perturbation theory:
\begin{equation}
 \ket{\psip_L} - \ket{\psiinf} \sim \ket{\psip^{(1)}_L}
 = \int_{0}^{\infty} \dd k\,\frac{1}{E_\infty+E_k}\sum_{l,{m}}
 \sqrt{\frac{2l+1}{4\pi} } a_{klm} \ket{\psi_{\infty,klm}} \,.
\label{eq:psiL-psiInf}
\end{equation}
The coefficients are
\begin{equation}
 a_{klm} = \sqrt{\frac{4\pi}{2l+1}}
 \mbraket{\psi_{\infty,klm}}{\Delta V_C}{\psiinf} \,,
\label{eq:a-klm}
\end{equation}
where we have pulled out the angular normalization factor in order to keep later
expressions as simple as possible.
It will turn out useful to furthermore split $\ket{\psip^{(1)}_L}$ as
\begin{multline}
 \ket{\psip^{(1)}_L} = \int_{0}^{\infty} \dd k\,\frac{1}{E_\infty+E_k}\sum_{l}
 \sqrt{\frac{2l+1}{4\pi} } a_{kl0} \ket{\psi_{\infty,kl0}}+
 \int_{0}^{\infty} \dd k\,\frac{1}{E_\infty+E_k}\sum_{l,{m\neq0}}
 \sqrt{\frac{2l+1}{4\pi} } a_{klm} \ket{\psi_{\infty,klm}} \\
 \equiv \ket{\psip^{(1)}_{L,m=0}} + \ket{\psip^{(1)}_{L,m\neq0}} \,.
 \label{eq:psip-diff}
\end{multline}
The contribution from $\ket{\psip^{(1)}_{L,m\neq0}}$ to $\Edp'_1(L)$ is
\begin{equation}
  \int_{B^2_{+z}}\dd x \dd y \frac{\partial}{\partial z}\left(
  {-}2\braket{\psiinf}{\vec r} \braket{\vec r}{\psip^{(1)}_{L,m\neq0}}
 \right)
 = \int_{0}^{\infty} \dd k\,\frac{1}{E_\infty+E_k}
 \sum_{l,m\neq 0}\sqrt{\frac{2l+1}{4\pi}}
 \int_{B^2_{+z}}\dd x \dd y\, a_{klm}\frac{\partial}{\partial z}
 \Big[\psiinf(r){\psi_{\infty,klm}(\vec r)}\Big] \,,
\end{equation}
and similar to Eq.~\eqref{eq:withsphericalsym-2} the
integration domain $B^2_{+z}$ essentially reduces to the disc region
$S$:
\begin{multline}
 \int_{B^2_{+z}}\dd x \dd y\, a_{klm}\frac{\partial}{\partial z}
 \Big[\psiinf(r){\psi_{\infty,klm}(\vec r)}\Big] \\
 = \int_{0}^{2\pi}\dd \phi\, \ee ^ {\ii m \phi}
 \int_0^{\pi/4}\dd \theta \,\sqrt{g(\theta)} \, a_{klm}
 \frac{\partial}{\partial z} \left[
  \psiinf(r)\sqrt{\frac{(2l+1)(l-m)!}{4\pi(l+m)!}}
  P^{m}_l(\cos\theta) \psi_{\infty,kl}(r)
 \right]
 + \OO(\ee^{{-}(\sqrt2/2+1/2)\kappa L}) \,.
\end{multline}
In the integral above, the only dependence on $\phi$ is through
$\ee^{\ii m \phi}$, so the leading term vanishes identically and the entire
contribution of
$\ket{\Delta\psip_{L,m\neq0}}$ to $\Edp'_1(L)$ is of higher exponential order.
Contributions from $\ket{\psip^{(1)}_{L,m\neq0}}$ to the second term,
$\Edp'_2(L)$, however, do not automatically vanish.
Therefore, we keep the following discussion of the coefficients $a_{klm}$
general and consider all $m$.

\medskip
Asymptotically, the scattering wave functions take the form
\begin{equation}
 \psi_{\infty,klm}(\vec r)
 = Y_l^{m}(\theta,\phi)\,\psi_{\infty,kl}(r)
 \stackrel{r\to\infty}{\sim} Y_l^{m}(\theta,\phi)\, \sqrt{\frac{2}{\pi}}
 \frac{1}{r} \sin\big[kr - \eta\log(2kr) + l\pi/2 +\delta_l(k)\big] \,,
\label{eq:psi-scatt-asympt}
\end{equation}
and we are allowed to assume this form in all expressions that follow because
$\Delta V_C$ vanishes in the central cube.
Introducing then complex radial functions
\begin{equation}
 \psi^C_{\infty,kl}(r)\equiv \sqrt{\frac{2}{\pi}}\frac{1}{r}
 \ee^{{-}\ii[kr - \eta_k\log(2kr) + l\pi/2 + \delta_l(k)]}
 = \ee^{{-}\ii[\delta_l(k)-\delta_0(k)+l\pi/2]}\psi^C_{\infty,k0}(r) \,,
\label{eq:psikc}
\end{equation}
we can define coefficients $a_{klm}^C$ exactly as in Eq.~\eqref{eq:a-klm}, only
with $\psi_{\infty,klm} \to \psi_{\infty,klm}^C$.
We have $\abs{a_{klm}} \leq \abs{a_{klm}^C}$ since $\psi_{\infty,klm} =
\Im\,\psi_{\infty,klm}^C$, asymptotically.
Since our mail goal is to derive bounds for $\abs{a_{klm}}$, we can work
in the following with the complex expressions, which simplifies the derivation.
Written out explicitly, we have
\begin{equation}
 a_{klm}^C =
 \int_{L/2}^{\infty} \dd^3r \,\sqrt{\frac{(l-m)!}{(l+m)!}} P_l^m(\cos\theta) \,
 \ee^{\ii m \phi}\psiinf(r) \Delta V_C(\vec{r}) \psi^C_{\infty,kl}(r) \,,
\end{equation}
where the radial integral only starts at $L/2$ owing to the limited support of
$\Delta V_C$.
It holds that $\abs{P_l(\cos\theta)}\leq 1$~\cite{Szego:1939}, and from Eq.~(5)
in Ref.~\cite{Lohoefer:1998aa} we have a similar inequality for
$P_l^m(\cos\theta)$:
\begin{equation}
 \abs{\sqrt{\frac{(l-m)!}{(l+m)!}}P^m_l(\cos\theta)\ee^{\ii m \phi}} \leq 1
 \label{eq:ineq1-assoc-legendre} \,.
\end{equation}
Since furthermore $\abs{\psi^C_{\infty,kl}(r)} \leq \sqrt{2/\pi}/r< 2/L$,
we can bound contributions to $a_{klm}^C$ from the region $r > L/\sqrt{2}$ as
follows:
\begin{multline}
 \abs{\int_{L/\sqrt{2}}^{\infty} \dd^3r \,
 \sqrt{\frac{(l-m)!}{(l+m)!}} P^m_l(\cos\theta) \,
 \psiinf(r) \Delta V_C(\vec{r}) \psi^C_{\infty,kl}(r)}
 \leq \int_{L/\sqrt{2}}^{\infty}\dd^3 r \abs{\psiinf(r)\Delta V_C(\vec r)\frac2L}
\\\leq \frac{2}{L} \left(6\psiinf(L/\sqrt{2}) \int_{B^3+\hat z L} \dd^3r
 \abs{\Delta V_C(\vec r)}
 + \sum_{\abs{\vec n}> 1} \psiinf(r_{\vecn}) \int_{B^3+\vec n L} \dd^3r
 \abs{\Delta V_C(\vec r)}\right)
 \,.
\label{eq:asymptotic-region-bound}
\end{multline}
In the last step above we have partitioned the entire space outside the central
cube into disjoint shifted boxes, and we use $r_{\vec n}$ to denote the
minimum distance of each shifted box $B^3+\vec n L$ to origin.
We can bound the expression this way because we consider an $S$-wave bound state
(no angular dependence in $\psiinf(r_{\vecn})$) and because the radial part is
monotonically decreasing in the asymptotic region.
As next step we use
\begin{equation}
 \int_{B^3+\vec n L}\dd^3r \abs{\Delta V_C(\vec r)}
 \leq \int_{B^3}\dd^3r \abs{ V_C(\vec r)}
\end{equation}
to remove the $\vec{n}$ dependence of the integral in the second term, which
reduces the problem to considering the sum
$\sum_{\abs{\vec n}>1} \psiinf(r_{\vec n})$.
Defining $S_k = \{\vec{n} : k L < \abs{\vec{n}} \leq (k+1) L\}$, we can
rearrange it as
\begin{equation}
 \sum_{\abs{\vec n}>1} \psiinf(r_{\vec n})
 = \sum_{k=1}^{\infty}\left(
  \sum_{\vec n \in S_k} \psiinf(r_{\vec n})
 \right) \,.
\end{equation}
There are 26 elements in $S_1$, and the smallest possible $\abs{r_\vecn}$ for
that set is $L/\sqrt{2}$.
Separating out those terms, we can write
\begin{equation}
 \sum_{\abs{\vec n}>1} \psiinf(r_{\vec n})
 \leq 26 \psiinf(L/\sqrt{2})
 + \sum_{k=2}^{\infty}\text{card}(S_k)\psiinf\big((k-1)L\big) \,,
\label{eq:psiinf-sum}
\end{equation}
and for the number of elements in each set $S_k$ it holds
that~\cite{Ivic:2004aa}
$\text{card}({S_k}) = 4 \pi (k+1)^2 +\OO(k^{\theta})$
with $1\leq\theta\leq2$.
That is sufficient for us to conclude that the sum
converges---$\psiinf\big((k-1)L\big)$ rapidly decreases with increasing
$k$)---and overall the first term in Eq.~\eqref{eq:psiinf-sum} dominates.
That term can be combined with the first term in
Eq.~\eqref{eq:asymptotic-region-bound}, and overall we have shown that
\begin{equation}
 \abs{\int_{L/\sqrt{2}}^{\infty} \dd^3r \,\sqrt{\frac{(l-m)!}{(l+m)!}}
  P^m_l(\cos\theta) \,\psiinf(r) \Delta V_C(\vec{r}) \psi^C_{\infty,kl}(r)
 } = \OO(\ee^{{-}\kappa L/\sqrt{2}}) \,.
\end{equation}
We can use this result to restrict the radial integral in the bound for
$a_{klm}^C$ to merely $[L/2,L/\sqrt{2}]$.
With that in mind we introduce now
\begin{equation}
 I_{klm} = \int_{-1}^{1} \dd\!\cos\theta\,\sqrt{\frac{(l-m)!}{(l+m)!}}
 P^m_l(\cos\theta)\Theta_{km}(\cos\theta)
\end{equation}
with
\begin{spliteq}
 \Theta_{km}(\cos\theta)
 &= \int_{L/2}^{L/\sqrt{2}} \dd r\,r^2
 \int_{0}^{2\pi}\dd\phi \, \cos(m\phi) \,
 \psi^C_{\infty,k0}(r) \Delta V_{C}(\vec r) \psiinf(r) \\
 & = \int_{L/2}^{L/\sqrt{2}} \dd r\,r^2 \int_{0}^{2\pi}\dd\phi \,
 \ee^{\ii m\phi}\psi^C_{\infty,k0}(r) \Delta V_{C}(\vec r) \psiinf(r) \,,
\label{eq:Theta-km}
\end{spliteq}
noting the reflection symmetry of $\psi^C_{\infty,k0}(r) \Delta V_{C}(\vec r)
\psiinf(r)$ around the $z=0$ plane.
With these definitions we have
\begin{equation}
 \abs{a_{klm}}\leq\abs{a_{klm}^C}
 \leq \abs{I_{klm}} + \OO(\ee^{{-}\kappa L/\sqrt{2}})
\label{eq:a_klm-ineq}
\end{equation}
and we therefore proceed to analyze $\abs{I_{klm}}$.

The relevant region where the integrand defining $\Theta_{km}(\cos\theta)$ does
not vanish is visualized in Fig.~\ref{fig:CubicCookie}.
Specifically, what is shown there is the ``cookie'' shaped volume sticking out
of the face $B^2_{{+}z}$, and there are five additional such shapes related to
the one that is shown by cubic symmetry.
We now furthermore introduce the splitting
\begin{equation}
 I_{klm} = I_{klm}^{\text{I}} + I_{klm}^{\text{II}} + I_{klm}^{\text{III}} \,,
\end{equation}
defined by $\cos\theta$ domains $[{-}1,{-}1/\sqrt{2}]$,
$[{-}1/\sqrt{2},1/\sqrt{2}]$, and $[1/\sqrt{2},1]$, respectively.
Since $\Delta V_{C}(\vec r)$ is an even function of $\cos\theta$, it follows
that $I_{klm}^{\text{I}} = I_{klm}^{\text{III}}$ and we therefore first
consider $I_{klm}^{\text{III}}$, corresponding to the domain shown explicitly in
Fig.~\ref{fig:CubicCookie}.
For that angular range we can use the properties of $\Delta V_C(\vecr)$ to
reduce the radial domain even further:
\begin{equation}
 \Theta_{km}(\cos\theta)
 = \int_{L/(2\cos\theta)}^{L/\sqrt{2}} r^2\dd r
 \int_{0}^{2\pi} \dd\phi \, \ee^{\ii m \phi}
 \psi^C_{\infty,k0}(r) \Delta V_{C}(\vec r) \psiinf(r)
 \mathtext{for} \frac{1}{\sqrt{2}} \leq \cos\theta \leq 1 \,.
\end{equation}
Within this region, $\Delta V_C(\vecr)$ is still invariant with respect to
rotations around $z$-axis, so there is no dependence on $\phi$ other than
through $\ee^{\ii m \phi}$, and from that it follows that
$I_{klm}^{\text{III}}=0$ for $m\neq0$.
Therefore, we need only keep $I_{kl0}^{\text{III}}$ in the following, and by
symmetry also only $I_{kl0}^{\text{I}}$.

\begin{figure}[tbhp] \centering
 \includegraphics[width=0.35\textwidth]{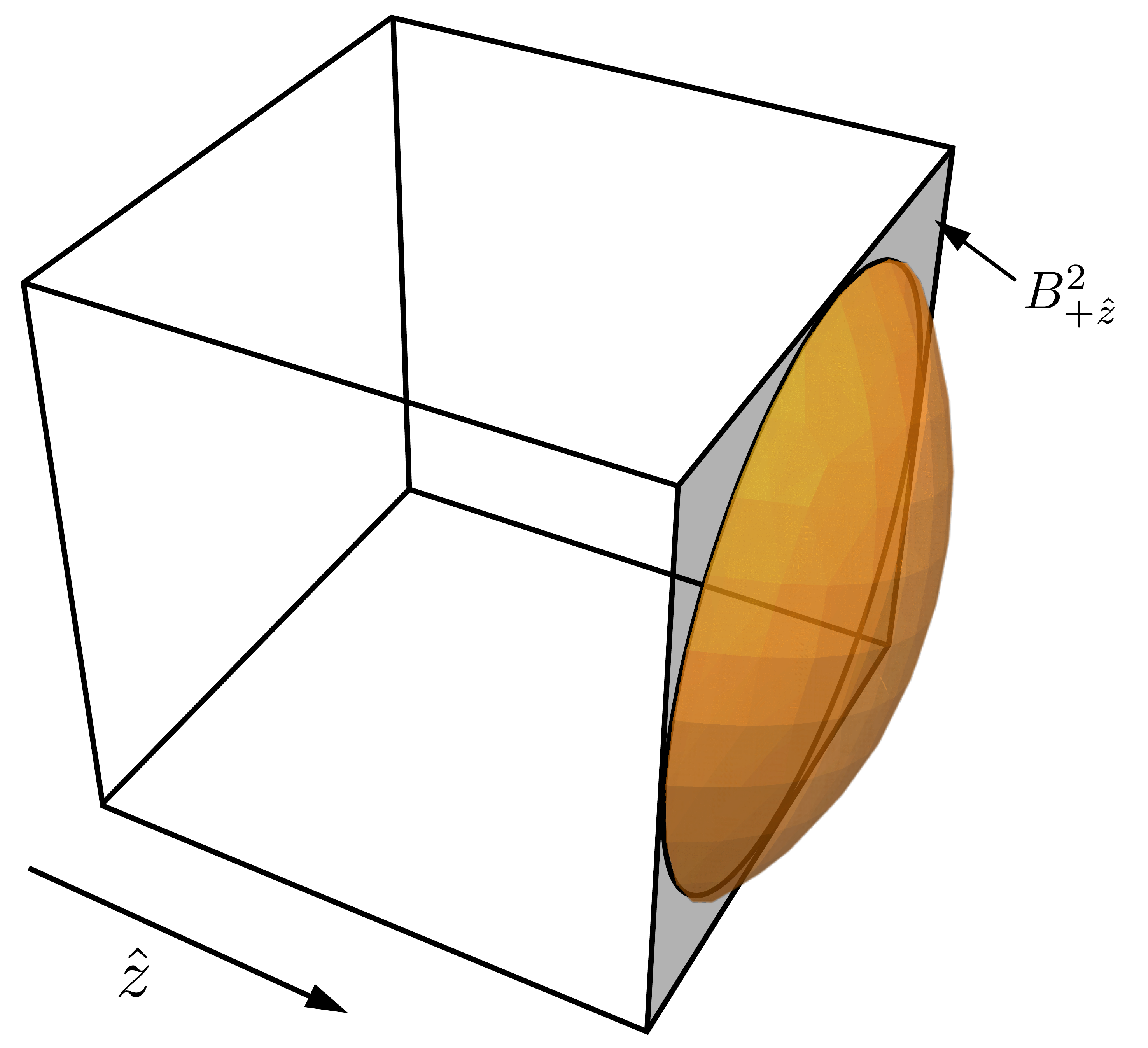}
 \caption{Visualization of the key integration region for the second correction
  term (see text).}
 \label{fig:CubicCookie}
\end{figure}

To study further the $m=0$ terms, we note that the key advantage of having
narrowed down the domains for both the radial integration as well as for
$\cos\theta$ is that we have are left with a simple expression for $\Delta
V_{C}(\vec r)$:
\begin{equation}
 \Delta V_{C}(\vec r) = \frac\gamma {2\mu}
 \left(\frac{1}{\abs{\vec{r}-\hat{z}L}}-\frac{1}{\abs{\vec{r}}}\right)
 \mathtext{for} \frac{L}{2} \leq \abs{\vec{r}} \leq \frac{L}{\sqrt{2}}
 \mathtext{,} \frac{1}{\sqrt{2}} \leq \cos\theta \leq 1 \,.
\end{equation}
This form we can expand around $r=L/(2\cos\theta)$ as
\begin{equation}
 \frac{1}{\abs{\vec{r}-\hat{z}L}}-\frac{1}{\abs{\vec{r}}}
 = 4\left(r-\frac L{2\cos\theta}\right)
 \frac{\cos^2\theta \, (1+\cos2\theta)}{L^2}
 + \OO\left(\frac{\big(r-L/(2\cos\theta)\big)^2}{L^3}\right) \,,
\end{equation}
and the result can be integrated over $\phi$ trivially.
The integration over $r$ can be performed using the asymptotic forms of the
bound-state and scattering wave functions~\cite{Olver:1965as}:
\begin{subalign}[eqs:psi-asympt]
\label{eq:psi-bound-asympt}
 \psiinf(r) &\stackrel{r\to\infty}{=}
 \frac{A_\kappa}{r} W_{-\bar\eta,1/2}(2\kappa r)
 \sim \frac{A_{\kappa}}{r}\frac{\ee^{-\kappa r}}{(2\kappa r)^{\bar\eta}} \,, \\
 \psi^C_{\infty,k0}(r) &\stackrel{r\to\infty}{=}
 \frac{A_{k0}}{r} \frac{\ee^{-\ii kr}}{(2kr)^{\ii \eta_k}} \,,
\end{subalign}
with $A_\kappa = {A_\infty}/{\sqrt{4\pi}}$ and
$A_{kl} =\sqrt{{2}/{\pi}}\,\ee^{{-}\ii[l\pi/2 + \delta_l(k)]}$.
We use these forms to carry out the integration and then we keep only the
leading terms for the result, noting that corrections are suppressed by inverse
powers of $\kappa L$.
In order to write the result in a simple way, we express the leading terms
again in terms of $\psiinf(r),\psi^C_{\infty,k0}(r)$,
with $r={L}/{(2\cos\theta)}$:
\begin{equation}
 \Theta_{k0}(\cos\theta) = \frac{\pi\gamma}{\mu} \,
 \psi^C_{\infty,k0}\!\left(\frac{L}{2\cos\theta}\right)
 \psiinf\!\left(\frac{L}{2\cos\theta}\right)
 \times \frac{(\cos\theta)^2}{(\kappa+\ii k)^2}
 \left(1+\OO\left(\frac{1}{\kappa L}\right)\right)
 \mathtext{as} \kappa L \to \infty \,.
\label{eq:Theta-k-L-asympt}
\end{equation}

With some foresight we define now
\begin{equation}
 \tilde{\Theta}_{k0}(\cos\theta) = \frac{\Theta_{k0}(\cos\theta)}{\Theta_{k0}(1)}
 \exp\left(\frac{(\kappa+\ii k)(1-\cos\theta)L}{2}\right)
\end{equation}
and consider the expansion of this function around $x \equiv \cos\theta = 1$,
\begin{equation}
 \tilde{\Theta}_{k0}(x) = \sum_{n=0}^{\infty} C_n (1-x)^n \,,
\label{eq:Theta-tilde-series}
\end{equation}
with $C_0=1$ by construction.
Using Eq.~\eqref{eq:Theta-k-L-asympt} and inserting the asymptotic forms of the
wave functions as given in Eqs.~\eqref{eqs:psi-asympt},
we obtain
\begin{equation}
 \Tilde{\Theta}_{k0}(x)
 = \exp\!\left({-}(\kappa +\ii k)L \frac{(1-x)^2}{2x}\right)
 x^{4+\bar\eta+\ii\eta_k}
 \times \left(1+\OO\left(\frac{x}{\kappa L}\right)\right) \,.
\end{equation}
From this expression we can explicitly calculate the coefficients $C_n$ in
Eq.~\eqref{eq:Theta-tilde-series} as
\begin{equation}
 C_n = \frac{1}{n!}\frac{\dd^n}{\dd^n x} \Tilde{\Theta}_{k0}(x)\bigg|_{x=1}
 = \sum_{n'=0}^n \left[
  (4+\bar\eta+\ii\eta_k-n'+1)_{n'} + \OO\left(\frac{1}{\kappa L}\right)
 \right] \binom{n}{n'}\frac{(n-n')!}{n!} D_{n-n'} \,,
\end{equation}
where we have used the Pochhammer symbol $(a)_i=a(a+1)\ldots(a+i-1)$ and
defined
\begin{equation}
 D_n = \frac{1}{n!}\frac{\dd^n}{\dd x^n}
 \ee^{{-}\varkappa(1-x)^2/x}\bigg|_{x=1} \,,
\end{equation}
with $\varkappa = {(\kappa +\ii k)L}/{2}$.
It is easy to check that $D_0=1, D_1=0, D_2={-}\varkappa$, and via the
recurrence relation
\begin{equation}
 (3+n) D_{n+3} + (2n+4) D_{n+2} + (1+n+2\varkappa) D_{n+1} + \varkappa D_n = 0
\end{equation}
it follows that for all $D_n$ it holds that
\begin{equation}
 D_n = \OO\big((\varkappa)^{\lfloor n/2\rfloor}\big)
 \mathtext{as} \abs{\varkappa} \to \infty \,.
\end{equation}
This then implies
\begin{equation}
 C_n = \OO((\kappa L)^{\lfloor n/2\rfloor})
 \mathtext{as}  \kappa L \to \infty \,.
\end{equation}
With this expansion and scaling for the coefficients, we can go back to
considering $I_{kl0}^{\text{III}}$.
Since the functions comprising $\Tilde{\Theta}_{k0}(x)$ are analytic on the
integration domain $1/\sqrt{2} \leq x \leq 1$, and because the power series for
$\Tilde{\Theta}_{k0}(x)$ converges absolutely, we are allowed to interchange
integration and summation to obtain
\begin{equation}
 I_{kl0}^{\text{III}} = \int_{1/\sqrt{2}}^1 \dd x \, P_l(x) {\Theta}_{k0}(x)
 = {\Theta}_{k0}(1) \times \sum_{n=0}^{\infty} C_n
 \int_{1/\sqrt{2}}^1 \dd x \, P_l(x) (1-x)^n
 \exp\!\left(\frac{(\kappa+\ii k)(1-x)L}{2}\right)
 \equiv \sum_{n=0}^{\infty} I_{kl0}^{\text{III},(n)} \,.
\label{eq:I-kl0-III}
\end{equation}
We consider first the leading term, $I_{kl0}^{\text{III},(n)}$.
Since
\begin{equation}
 \abs{\int_{-1}^{1/\sqrt{2}} \dd x \, P_l(x) \ee^{-(\kappa+\ii k)L(1-x)/2}}
 \leq \int_{-1}^{1/\sqrt{2}} \dd x \, \abs{P_l(x) \ee^{-(\kappa+\ii k)L(1-x)/2}}
 = \OO\big(\ee^{{-}(1-1/\sqrt{2})\kappa L}\big)
\end{equation}
comes with additional exponential suppression in addition to the leading
behavior $\OO\big(\ee^{{-}\kappa L/2}\big)$ from ${\Theta}_{k0}(1)$,
we can actually extend the integration domain for $x$ back to $[{-}1,1]$.
Using then
\begin{equation}
 \int_{-1}^{1} \dd x\, P_l(x)\ee^{{-}\varkappa(1-x)}
 = \sqrt{\frac{2\pi}{a}}\mathcal{I}_{l+1/2}(\varkappa)\ee^{{-}\varkappa} \,,
\label{eq:Fl-0}
\end{equation}
where $\mathcal{I}_{l+1/2}$ is a modified Bessel function,
we obtain
\begin{equation}
 \abs{I_{kl0}^{\text{III},(0)}} = \abs{{\Theta}_{k0}(1)}
 \times \abs{
  \sqrt{\frac{2\pi}{\varkappa}}
  \mathcal{I}_{l+1/2}(\varkappa)
  \ee^{{-}\varkappa}
 } \mathtext{with} \varkappa = {(\kappa +\ii k)L}/{2} \,.
\label{eq:I-kl0-III-0}
\end{equation}
For the remaining terms with $n>0$ we can similarly extend the integration
domain to $[-1,1]$ and are then faced with integrals of the form
\begin{equation}
 F_{l}^{(n)}(\varkappa)
 = \int_{-1}^{1} \dd x\, (1-x)^n P_l(x) \ee^{{-}\varkappa(1-x)} \,.
\end{equation}
$F_{l}^{(0)}(\varkappa)$ is exactly the expression in Eq.~\eqref{eq:Fl-0},
and the general case we can new obtain via
\begin{equation}
 F_{l}^{(n)}(\varkappa)
 = ({-}1)^n \frac{\dd^n}{\dd\varkappa^n} F_{l}^{(0)}(\varkappa) \,,
\end{equation}
by differentiation under the integral.
A straightforward inductive argument shows that
$F_{l}^{(n)}(\varkappa) = \OO(1/\varkappa^n) \times F_{l}^{(0)}(\varkappa)$.
What we have ultimately learned from this analysis is that each factor $(1-x)$
in Eq.~\eqref{eq:I-kl0-III} brings a suppression of $1/(\kappa L)$ after
integration, which means that we are permitted to keep only the leading term,
given by Eq.~\eqref{eq:I-kl0-III-0}.

For the overall prefactor we find that
\begin{equation}
\label{eq:Theta-k-1}
 \abs{{\Theta}_{k0}(1)}
 = \frac{8\pi}{\mu}\frac{\kappa^2}{\kappa^2+k^2}
 \frac{\bar\eta}{\kappa L} \psiinf(r_0)
 \times\left[1+\OO\left(\frac{1}{\kappa L}\right)\right] \,,
\end{equation}
and with the asymptotic form
\begin{equation}
 \sqrt{\frac{2\pi}{\varkappa}}\mathcal{I}_\ell(\varkappa)\ee^{{-}\varkappa}
 = \frac{1}{\varkappa} \exp\!\left({-}\frac{\ell^2}{2\varkappa}\right)
 \left[1+\OO\left(\frac{1}{\kappa L}\right)\right]
 \mathtext{for}
 \abs{\varkappa} \to \infty,\ \ell \ll \abs{\varkappa}
\label{eq:I-nu-asympt}
\end{equation}
we finally arrive at
\begin{equation}
 \abs{I_{kl0}^{\text{III}}}
 = \frac{8\pi}{\mu}\frac{\kappa^2}{\kappa^2+k^2}
 \frac{\bar\eta}{\kappa L} \psiinf(r_0)
 \abs{
  \frac{\ee^{-\frac{(l+1/2)^2}{(\kappa+\ii k) L}}}{(\kappa+\ii k) L}
 } \times \left[1+\OO\left(\frac{1}{\kappa L}\right)\right] \,.
\label{eq:I-kl0-III-final}
\end{equation}
As noted before, $I^{\text{III}}_{klm} = 0$ for $m\neq0$.

\medskip
We now turn to analyzing the contribution $\abs{I_{klm}^{\text{II}}}$,
corresponding to the angular range
${-}1/\sqrt{2} \leq \cos\theta \leq 1/\sqrt{2}$.
The integration domain for this term is comprised of four cookie shapes as in
Fig.~\ref{fig:CubicCookie}, sticking out of the faces $B^2_{{+}x}$ and
$B^2_{{+}y}$.
Since we are considering an $S$-wave bound state in infinite volume that is the
ground state of the two-particle system, it is valid to assume that $\psip_L$
in the box falls into the $A_1^+$ representation of the cubic group, with the
superscript indicating positive parity.
Parity implies that $l$ needs to be even in Eq.~\eqref{eq:psiL-psiInf}, and
since we have (in particular) invariance under rotations by $\pi/2$ around the
$z$ axis, only $m \equiv 0\ \text{mod}\ 4$ can contribute to
$\ket{\tilde{\psi}_L^{(1)}}$~\cite{Muggli:1972aa}.
Therefore, all four parts of the integration domain will give equal
contributions.
Among these terms we can then focus on $B^2_{{+}x}$, where we have
${-}\pi/4 \leq \phi \leq \pi/4$ and essentially we can follow the same steps as
before, with only one major change: the boundary for the radial integral is now
defined by $r={L}/{(2\sin\theta\cos\phi)}$.
This means we need to evaluate the azimuthal integral explicitly instead of
simply picking up a factor $2\pi$.
We obtain:
\begin{equation}
\Theta_{km}(\cos\theta)
 = \frac{4\pi\gamma}{\mu} \int_{{-}\pi/4}^{\pi/4} \dd\phi\,\ee^{\ii m\phi} \,
 \psi^C_{\infty,k0}\!\left(\frac{L}{2\sin\theta\cos\phi}\right)
 \psiinf\!\left(\frac{L}{2\sin\theta\cos\phi}\right)
 \times \frac{(\sin\theta\cos\phi)^2}{(\kappa+\ii k)^2}
 \left(1+\OO\left(\frac{1}{\kappa L}\right)\right)
\label{eq:Theta-k-L-asympt-II}
\end{equation}
as $\kappa L \to \infty$ in $I_{klm}^{\text{II}}$.
We define
\begin{equation}
 \varkappa' = \frac{\kappa+\ii k}{2\sin\theta} \,,
\end{equation}
and after a few steps---using again the exponential asymptotic forms of both
wave functions appearing in Eq.~\eqref{eq:Theta-k-L-asympt-II}---we can rewrite
$\Theta_{km}(\cos\theta)$ as
\begin{multline}
 \Theta_{km}(\cos\theta)
 = \frac{4\pi\gamma}{\mu}
  \psi^C_{\infty,k0}\!\left(\frac L{2\sin\theta}\right)
  \psiinf\!\left(\frac L{2\sin\theta}\right)
  \frac{(\sin\theta)^2}{(\kappa+\ii k)^2}
  \left(1+\OO\left(\frac{1}{\kappa L}\right)\right) \\
 \null \times \int_{{{-}\pi/4}}^{{\pi/4}} \dd \phi\,
  \ee^{-\frac{\varkappa'}{2\cos\phi}
  + \frac{\varkappa'}{2}+\ii m\phi} (\cos\phi)^{2+\bar\eta+\eta_k} \,.
\end{multline}
Since $\abs{\varkappa'} \sim \kappa L $ is large, we can then use the saddle
point approximation method to carry out the integral over $\phi$:
\begin{multline}
 \int_{{{-}\pi/4}}^{{\pi/4}} \dd \phi \,
  \ee^{{-}\frac{\varkappa'}{2\cos\phi} + \frac{\varkappa'}{2}+\ii m \phi}
  (\cos \phi)^{2+\bar\eta+\eta_k}
 = \int_{{{-}\pi/4}}^{{\pi/4}}\dd \phi \,
  \ee^{
   {-}\frac{\varkappa'}{4} \phi^2 -(\eta_k+\bar\eta+2)\phi^2/2 +\ii m \phi
   + \OO(\phi^4)
  } \\
 = \sqrt{\frac{2\pi}{\varkappa'/2+\eta_k+\bar\eta+2}}
 \ee^{-\frac{m^2}{4(\varkappa'/2+\eta_k+\bar\eta+2)}}
  \left(1+\OO\left(\frac{1}{\kappa L}\right)\right)
 =\sqrt{\frac{2\pi }{\varkappa'}}
 \ee^{-\frac{m^2}{2\varkappa}}
  \left(1+\OO\left(\frac{1}{\kappa L}\right)\right) \,.
\end{multline}
With that we obtain
\begin{equation}
 \Theta_{km}(\cos\theta)
 = \frac{8\gamma}{\mu}\sqrt{\frac{4\pi\sin\theta}{(\kappa+\ii k) L}}
 \ee^{-\frac{m^2}{(\kappa+\ii k) L}}
 \psi^C_{\infty,k0}\!\left(\frac L{2\sin\theta}\right)
 \psiinf\!\left(\frac L{2\sin\theta}\right)
 \frac{(\sin\theta)^2}{(\kappa+\ii k)^2}
 \left(1+\OO\left(\frac{1}{\kappa L}\right)\right) \,.
\end{equation}
Although the angular dependence now exclusively appears via $\sin\theta$, we
will keep writing the argument of $\Theta_{km}$ as $\cos\theta$, noting that in
principle the entire expression could be rewritten using $\sin\theta =
\sqrt{1-\cos^2\theta}$.
With that in mind we introduce $y=\sin\theta$ and proceed now similar to what
we did following Eq.~\eqref{eq:Theta-k-L-asympt}, except that now we consider an
expansion around $y = 1$.
Since $y = \sin\theta = 1$ implies $x = \cos\theta=0$, we furthermore introduce
the auxiliary function
\begin{equation}
 \Theta^{(i)}_{km}(\sin\theta)
 = \frac{\Theta_{km}(\cos\theta)}{\Theta_{km}(0)}
 \ee^{{-}\varkappa(1-\sin\theta)}
 (\sin\theta)^{1/2-i-j_m} \mathtext{,}
 j_m = \begin{cases}
  1/2 & m = 0 \\
  m   & m \neq 0
 \end{cases} \,,
\end{equation}
with integers $i\geq0$ and using $\varkappa = (\kappa + \ii k)L/2$ as before.
The purpose of including the additional factors $(\sin\theta)^{{-}i-j_m}$ will
become clear in the following.
We can expand $\Theta^{(i)}_k(y)$ around $y=1$,
\begin{equation}
 \Theta^{(i)}_{km}(y)
 = \ee^{{-}\frac{\varkappa(1-y)^2}{y}}
 y^{5+{\eta} + \ii\eta_k - i-j_m}
 \times \left(1+\OO\left(\frac{1}{\kappa L}\right)\right)
 = \sum_{n=0}^{\infty} C_{m,n}^{(i)} (1-y)^n \,,
\end{equation}
as before by analyticity of all involved functions.
Similarly to what we did for $I_{klm}^{\text{III}}$, we find that
$C_{m,0}^{(i)}=1$ and
$C_{m,n}^{(i)} = \OO\big((\kappa L)^{\lfloor n/2\rfloor}\big)$ for $n \geq 1$,
independent of $m,i$.
Overall we have therefore established that
\begin{equation}
 \Theta_{km}(\cos\theta) = \Theta_{km}(0) \, (\sin\theta)^{i+j_m-1/2}
 \ee^{{-}\varkappa(1-\sin\theta)}
 \times \sum _{n=0}^{\infty}  C_{m,n}^{(i)} (1-\sin\theta)^n
\label{eq:sideTheta}
\end{equation}
on the domain $\pi/4 \leq \theta \leq 3\pi/4$, for any integer $i\geq0$,
and this form we can insert into
\begin{equation}
 I_{klm}^{\text{II}}
 = \int_{-1/\sqrt{2}}^{1/\sqrt{2}} \dd x \,\sqrt{\frac{(l-m)!}{(l+m)!}}
  P_l^m(x) {\Theta}_{km}(x) \,.
\label{eq:I-kl0-II-a}
\end{equation}
For even $m$ it holds that
\begin{equation}
  \sqrt{\frac{(l-m)!}{(l+m)!}}P^m_l(\cos\theta)  =
  \sqrt{\frac{(l+m)!}{(l-m)!}}P^{-m}_l(\cos\theta) \,,
\end{equation}
and therefore, since as mentioned before only $m$ that are multiples of $4$
can contribute, in the following we consider $m\geq 0$ only.
As is stated before, with an $A_1^+$ states we need only consider even $l$.
We can therefore use $\cos(\theta-l\pi/2) = ({-}1)^{l/2} \cos(\theta)$
to combine Theorem 8.21.5 and Eq.~8.4.13 in Ref.~\cite{Szego:1939} as follows:
\begin{equation}
 \sqrt{\frac{(l-m)!}{(l+m)!}}P^m_l(\cos\theta)
 = ({-}1)^l\frac{4}{\pi}\sum_{i=0}^{\infty}
 h_{lm,i}\frac{\cos\left[(l+j'_{m,i})(\theta-\frac\pi2)\right]}
 {(\sin\theta)^{i+j_m}}
 \mathtext{,} \pi/6 < \theta < 5\pi/6 \,,
\end{equation}
where
\begin{equation}
 j'_{m,i}= \begin{cases}
  i+1/2 &,\ m= 0 \\
  m-i   &,\ m> 0 \,,
 \end{cases}
\end{equation}
and
\begin{equation}
 h_{lm,i} = \begin{cases}
  \frac{(2l)!!(2i-1)!!\left(\frac12\right)_{i}}
  {2^{\frac12+i}(2l+1)!!(2i-1)!!\left(l+\frac32\right)_{i}}
  &,\ m= 0 \\
  \sqrt{\frac{(l-m)!}{(l+m)!}}
  \frac{(m)_l(m)_i(m-i)_i}{2^{m+i}i!l!(l+m-i)_i}
  &,\ m>0\ \text{and}\ i\leq m-1 \\
  0
  &,\ m>0\ \text{and}\ i> m-1 \,.
 \end{cases}
\label{eq:h-li}
\end{equation}
From above definitions it is obvious that $h_{lm,i}\geq0$, and
$j'_{m,i} \geq 1/2$ for $h_{lm,i}\neq 0$.
Switching back to using $\theta$ instead of $x$ to write the integral, we now
have:
\begin{equation}
 I_{klm}^{\text{II}}
 = \Theta_{km}(0) \times \sum_{n=0}^{\infty}\sum_{i=0}^{\infty}
 C^{(i)}_{nm} h_{lm,i}\times \frac{4}{\pi}\int_{\pi/4}^{3\pi/4} \dd\theta \,
 \cos\left[(l+j'_{m,i})(\theta-\frac\pi2)\right]
 (1-\sin\theta)^n \ee^{{-}\varkappa(1-\sin\theta)} \,.
\label{eq:I-kl0-II-b}
\end{equation}
A straightforward calculation gives that
\begin{equation}
 \frac{\Theta_{km}(0)}{\Theta_{k0}(1)}
 = \sqrt{\frac{16}{\pi(\kappa+\ii k) L}}\ee^{-\frac{m^2}{(\kappa+\ii k) L}}
 = \sqrt{\frac{8}{\pi\varkappa}}\ee^{-\frac{m^2}{2\varkappa}} \,,
\end{equation}
where we use $\Theta_{k0}(1)$ as reference because $\Theta_{km}(1)$ vanishes for
$m\neq 0$.
With arguments analogous to those for $I_{klm}^{\text{III}}$ with $m=0$,
we need only keep the $n=0$ term in Eq.~\eqref{eq:I-kl0-II-b} because all terms
with $n\geq1$ are suppressed by powers of $1/(\kappa L)$.
After a few steps we arrive at
\begin{equation}
 \abs{I_{klm}^{\text{II,(0)}}} = 16\sqrt{\frac{2}{\pi^3}}\abs{
 \frac{\Theta_{k0}(1)}{\sqrt{\varkappa}}\ee^{-\frac{m^2}{2\varkappa}}
 \sum_{i=0}^\infty h_{lm,i}
 \int_0^{\pi/4} \dd t \cos\left[\left(l+j'_{m,i}\right)t\right]
 ({-}1)^{l/2} \ee^{{-}\varkappa(1-\cos t)}
 } \times \left(1+\OO\left(\frac{1}{\kappa L}\right)\right) \,,
\label{eq:I-kl0-II-t}
\end{equation}
where we have changed the integration variable to $t = \theta - \pi/2$.
Note that we can drop the pure phase $({-}1)^{l/2}$ within the absolute value
in the following.
Then, since contributions from larger $t$ are suppressed, \ie,
\begin{equation}
 \abs{
 \frac{\Theta_{k0}(1)}{\sqrt{\varkappa}}\ee^{-\frac{m^2}{2\varkappa}}
 \int_{\pi/4}^\pi \dd t \cos\left[\left(l+j'_{m,i}\right)t\right]
 \ee^{{-}\varkappa(1-\cos t)}
 } = \OO\left(\ee^{{-}(1-1/(2\sqrt{2})){\kappa L}}\right) \,,
\end{equation}
we can extent the integration range in Eq.~\eqref{eq:I-kl0-II-t} from $0$ to
$\pi$.
For $m=0$, the infinite sum over $i$ does not spoil this argument because the
coefficients $h_{l,i}$ decrease rapidly with $i$, cf.~Eq.~\eqref{eq:h-li}.
Specifically, it holds that~\cite{Szego:1939}
\begin{equation}
 \frac{4}{\pi}\sum_{i=0}^{\infty}h_{lm,i} =
 ({-}1)^l \sqrt{\frac{(l-m)!}{(l+m)!}} P_l(0) \,.
\label{eq:hlm-sum}
\end{equation}
With the integration range extended in this way, we can again evaluate the
integral in terms of modified Bessel functions, similar to what we did for
$I_{kl0}^{\text{III}}$.
The key to that is the following relation (Eq.~(9.6.26) in
Ref.~\cite{Abramowitz:1964}):
\begin{equation}
 \mathcal{I}_{v}(\varkappa)
 = \frac{1}{\pi}\int_{0}^{\pi} \dd t \, \cos(vt) \ee^{\varkappa\cos t}
 - \frac{\sin v\pi}{\pi}
 \int_{0}^{\infty} \dd t \, \ee^{{-}\varkappa\cosh t -v t } \,.
\label{eq:I-nu-exp}
\end{equation}
With $v=l+j'_{m,i}>0$ and $\cosh t \geq 1$ for $t\in[0,\infty)$ it is clear that
the second in Eq.~\eqref{eq:I-nu-exp} can be neglected as
$\OO(\ee^{{-}\varkappa})$, and we obtain
\begin{equation}
 \abs{I_{klm}^{\text{II,(0)}}} = \frac{16}{\pi}\abs{
 \Theta_{k0}(1) \ee^{-\frac{m^2}{2\varkappa}}\, \ee^{{-}\varkappa}
 \sum_{i=0}^\infty h_{lm,i}
 \sqrt{\frac{2\pi}{\varkappa}} \mathcal{I}_{l+j'_{m,i}}(\varkappa)}
 \times \left[1+\OO\left(\frac{1}{\kappa L}\right)\right]
 + \OO\left(\ee^{{-}(1-1/(2\sqrt{2})){\kappa L}}\right) \,.
\end{equation}
We can then use Eq.~\eqref{eq:I-nu-asympt} again to write,
for $\abs{\varkappa} \sim \kappa L \to \infty$,
\begin{spliteq}
 \abs{I_{klm}^{\text{II,(0)}}}
 &\leq 4\abs{\Theta_{k0}(1)} \abs{\frac{4}{\pi}\sum_{i=0}^\infty h_{lm,i}}
 \abs{
  \frac{\ee^{-\frac{(l+1/2)^2+m^2}{(\kappa+\ii k) L}}}{(\kappa+\ii k) L}
 }\left[1+\OO\left(\frac{1}{\kappa L}\right)\right]
 + \OO\left(\ee^{{-}(1-1/(2\sqrt{2})){\kappa L}}\right) \\
 &= 4\abs{\Theta_{k0}(1)}\abs{\sqrt{\frac{(l-m)!}{(l+m)!}}P^m _l(0)}\abs{
  \frac{\ee^{-\frac{(l+1/2)^2+m^2}{(\kappa+\ii k) L}}}{(\kappa+\ii k) L}
 } \left[1+\OO\left(\frac{1}{\kappa L}\right)\right]
 + \OO\left(\ee^{{-}(1-1/(2\sqrt{2})){\kappa L}}\right)\,,
\label{eq:I-klm-II-0-final}
\end{spliteq}
where in the last step we have used Eq.~\eqref{eq:hlm-sum}.

\medskip
At this point we can now finally collect all results and conclude the argument.
For $m=0$, comparing the equation above with Eqs.~\eqref{eq:Theta-k-1}
and~\eqref{eq:I-kl0-III-final} gives
\begin{equation}
 \abs{I_{kl0}^{\text{II}}}
 \leq 4 \abs{P_l(0)} \abs{I_{kl0}^{\text{III}}}
 \times \left[1+\OO\left(\frac{1}{\kappa L}\right)\right] \,.
\end{equation}
Remembering that by symmetry $\abs{I_{kl0}^{\text{I}}} =
\abs{I_{kl0}^{\text{III}}}$, we ultimately have $\abs{a_{kl0}} \leq
6\abs{I_{kl0}^{\text{III}}}$, and inserting the expression for
$\abs{I_{kl0}^{\text{III}}}$ from Eq.~\eqref{eq:I-kl0-III-final} then gives
\begin{equation}
 \abs{a_{kl0}} \leq \frac{48\pi}{\mu}\frac{\kappa^2}{\kappa^2+k^2}
 \frac{\bar\eta}{\kappa L} \psiinf(r_0)
 \abs{
  \frac{\ee^{-\frac{(l+1/2)^2}{(\kappa+\ii k) L}}}{(\kappa+\ii k) L}
 } \times \left[1+\OO\left(\frac{1}{\kappa L}\right)\right] \,.
 \label{eq:a_kl0_finalineq}
\end{equation}
This now allows us to estimate first the perturbative correction stemming from
$\braket{\vec r}{\psip^{(1)}_{L,m=0}}$.
Keeping in mind that we are dealing with a surface integral over $B_{+z}^2$, we
have
\begin{equation}
 \abs{\braket{\vec r}{\psi_{\infty,kl0}}}
 \leq \sqrt{\frac{2l+1}{4\pi}}\abs{\psi_{\infty,kl}(r)}
 \leq \sqrt{\frac{2l+1}{2\pi^2}}\frac{1}{r_0} \,,
\end{equation}
and thus
\begin{spliteq}
 \abs{\braket{\vec r}{\psip^{(1)}_{L,m=0}}}
 &\leq \int_{0}^{\infty} \dd k\,\frac{1}{E_\infty+E_k}\sum_{l}
 \sqrt{\frac{2l+1}{4\pi} } \abs{\strut
  a_{kl0} \braket{\vec r}{\psi_{\infty,kl0}}
 } \\
 &= \frac{48\pi}{\mu}\frac{\bar\eta}{\kappa L}
 \int_{0}^{\infty} \dd k\,\frac{E_\infty}{(E_\infty+E_k)^2}
 \frac{\psiinf(r_0)}{4\pi r_0} \times
 \sum_{l} (2l+1) \abs{
  \frac{\ee^{-\frac{(l+1/2)^2}{(\kappa+\ii k) L }}}{(\kappa+\ii k)L}
 } \times \left[1+\OO\left(\frac{1}{\kappa L}\right)\right] \,.
\label{eq:psip-1-m0-sum}
\end{spliteq}
We now need to evaluate the sum over $l$.
If we denote
\begin{equation}
 f(t) = 2t \abs{
  \frac{\ee^{{-}\frac{t^2}{(\kappa+\ii k) L }}}{(\kappa+\ii k)L}
 } \,,
\end{equation}
then $f(t) \geq 0$ for $t \geq 0$ and the sum in Eq.~\eqref{eq:psip-1-m0-sum}
can be written as $\sum_{l=0}^\infty f(l+1/2)$.
It furthermore holds that
$\inf_{t\in [l,l+1]} f(t) \leq f(l+1/2) \leq \sup_{t\in [l,l+1]} f(t)$,
and
\begin{subalign}
  \sum_{l=0}^\infty \inf_{t\in [l,l+1]} f(t)
 \leq \sum_{l=0}^\infty f(l+1/2)
 \leq \sum_{l=0}^\infty \sup_{t\in [l,l+1]} f(t) \,,\\
 \sum_{l=0}^\infty \inf_{t\in [l,l+1]} f(t)
 \leq \int_0^\infty \dd t \, f(t)
 \leq \sum_{l=0}^\infty \sup_{t\in [l,l+1]} f(t) \,,
\label{eq:fsum-0}
\end{subalign}
where we applied the method of Darboux sums as explained in
Ref.~\cite{Royden:1988l}.
The unique maximum of $f(t)$ is found at
$t_0 = \sqrt{\abs{(\kappa+\ii k)L/2}}$, and with that we obtain
\begin{equation}
 \abs{
  \sum_{l=0}^\infty f(l+1/2) - \int_0^\infty \dd t f(t)
 } \leq \sum_{l=0}^\infty \left(
  \sup_{t\in [l,l+1]} f(t) - \inf_{t\in [l,l+1]} f(t)
 \right)
 \leq 2 \max_{t\in[0,\infty)}{f(t)}
 = 2f(t_0) = \OO\left(\frac{1}{(\kappa L)^{1/2}}\right) \,.
\label{eq:fsum-1}
\end{equation}
From this it follows that
\begin{equation}
 \sum_{l=0}^\infty f(l+1/2)
 = \int_0^\infty \dd t \, f(t) + \OO\left(\frac{1}{(\kappa L)^{1/2}}\right)
 = 1 + \OO\left(\frac{1}{(\kappa L)^{1/2}}\right) \,,
\label{eq:fsum-2}
\end{equation}
and therefore:
\begin{equation}
 \abs{\braket{\vec r}{\psip^{(1)}_{L,m=0}}}
 \leq \frac{48\pi}{\mu}\frac{\bar\eta}{\kappa L}
 \int_{0}^{\infty} \dd k\,\frac{E_\infty}{(E_\infty+E_k)^2}
 \frac{\psiinf(r_0)}{4\pi r_0}
 \times \left[1+\OO\left(\frac{1}{\kappa L}\right)\right]
 = \OO\left(\frac{\psiinf(r_0)}{(\kappa L)^{2}}\right) \,.
 \label{eq:delta-tildaphi-0}
\end{equation}
For $a_{klm}$ with $m \neq 0$ we have
$\abs{a_{klm}} \leq 2\abs{I_{klm}^{\text{III}}}$.
As noted below Eq.~\eqref{eq:I-kl0-III-final}, the first term vanishes for $m
\neq 0$, and for the second term we can use Eq.~\eqref{eq:I-klm-II-0-final}
together with Eq.~\eqref{eq:Theta-k-1}, giving
\begin{equation}
 \abs{a_{klm}}
 \leq \frac{8\pi}{\mu}\frac{\kappa^2}{\kappa^2 + k^2}\frac{\bar\eta}{\kappa L}
 \psiinf(r_0)
 \sqrt{\frac{(l-m)!}{(l+m)!}} P^m_l(0)\abs{
  \frac{\ee^{-\frac{(l+1/2)^2}{(\kappa+\ii k) L }}}{(\kappa+\ii k) L}
 } \times \left[1+\OO\left(\frac{1}{\kappa L}\right)\right] \,.
\end{equation}
To simplify this further, we use the inequality
\begin{equation}
 \frac{\sqrt{(2n)!}}{(2n)!!}
 = \sqrt{\frac{(2n-1)!!}{(2n)!!}}
 = \sqrt{\frac{\Gamma(n+1/2)}{\Gamma(1/2)\Gamma(n+1)}}
 < \left(\frac{1}{n\pi}\right)^{1/4}< {(n+1)}^{-1/4}
\label{eq:fac-bound}
\end{equation}
with $2n = l-m$ to obtain
\begin{equation}
 \sqrt{\frac{(l-m)!}{(l+m)!}} P^m_l(0)
 = \frac{\sqrt{(l-m)!(l+m)!}}{(l-m)!!(l+m)!!}
 < \left(\frac{{4}}{ (l-m+2)(l+m+2)}\right)^{1/4} \,.
\end{equation}
For $n = 0$, Eq.~\eqref{eq:fac-bound} is trivially true, and for $n>0$ it
follows as a special case of the Gautschi inequality for the Gamma function,
see for example Ref.~\cite{Wendel:1948aa}.
For general argument, from Eq.~\eqref{eq:ineq1-assoc-legendre} it follows
that
\begin{equation}
 \abs{Y_m^l(\theta,\phi)} \leq \sqrt{\frac{2l+1}{4\pi}}
 \mathtext{for} l \geq 1 \mathtext{and} \abs{m} \geq 1 \,,
\end{equation}
and therefore
\begin{equation}
 \abs{\braket{\vec r}{\psi_{\infty,klm}}}
 \leq\sqrt{\frac{2l+1}{4\pi}}\abs{\psi_{\infty,kl}(r)} \,.
\end{equation}
Recalling now the definition from Eq.~\eqref{eq:psip-diff}, we can see that
\begin{equation}
 \braket{\vec r}{\psip^{(1)}_{L,m\neq0}}
 = \int_{0}^{\infty} \dd k\,\frac{1}{E_\infty+E_k}\sum_{l,{m\neq0}}
 \sqrt{\frac{2l+1}{4\pi} } a_{klm} \braket{\vec r}{\psi_{\infty,klm}}
 = \OO\left(\frac{\psiinf(r_0)}{(\kappa L)^{2-1/4}}\right) \,.
\label{eq:psip-m-neq-0}
\end{equation}

The relative suppression of $\psip^{(1)}_{L}$, and therefore of $\psip_{L}$
can be checked with numerical calculations.
To that end, we use the lattice discretization method described in the main text
and consider an (arbitrarily chosen) sample scenario, using $\gamma = 1.0$ for
the Gauss-regulated Coulomb potential and an attractive short-range Gaussian
potential with $R=1$ and $V_0  = {-}6.2148$, with all quantities in units as
described in the main text.
We explicitly calculate the exact wave function $\psip_L$ at volume $L$ and we
furthermore approximate the infinite-volume wave function $\psiinf$ from a
lattice calculation in a volume larger than the range of interest.
With those wave functions at hand, we can calculate the relative difference
evaluated at $\vec{r} = \vec{r}_0$, which we show in Fig.~\ref{fig:plotwave}.
From the inverse square and inverse cubic guiding lines shown in the figure we
conclude that the numerical calculation is consistent with the relative
suppression of $\psip^{(1)}_{L}$ derived above.

\begin{figure}[tbhp]
 \centering
 \includegraphics[width=0.6\textwidth]{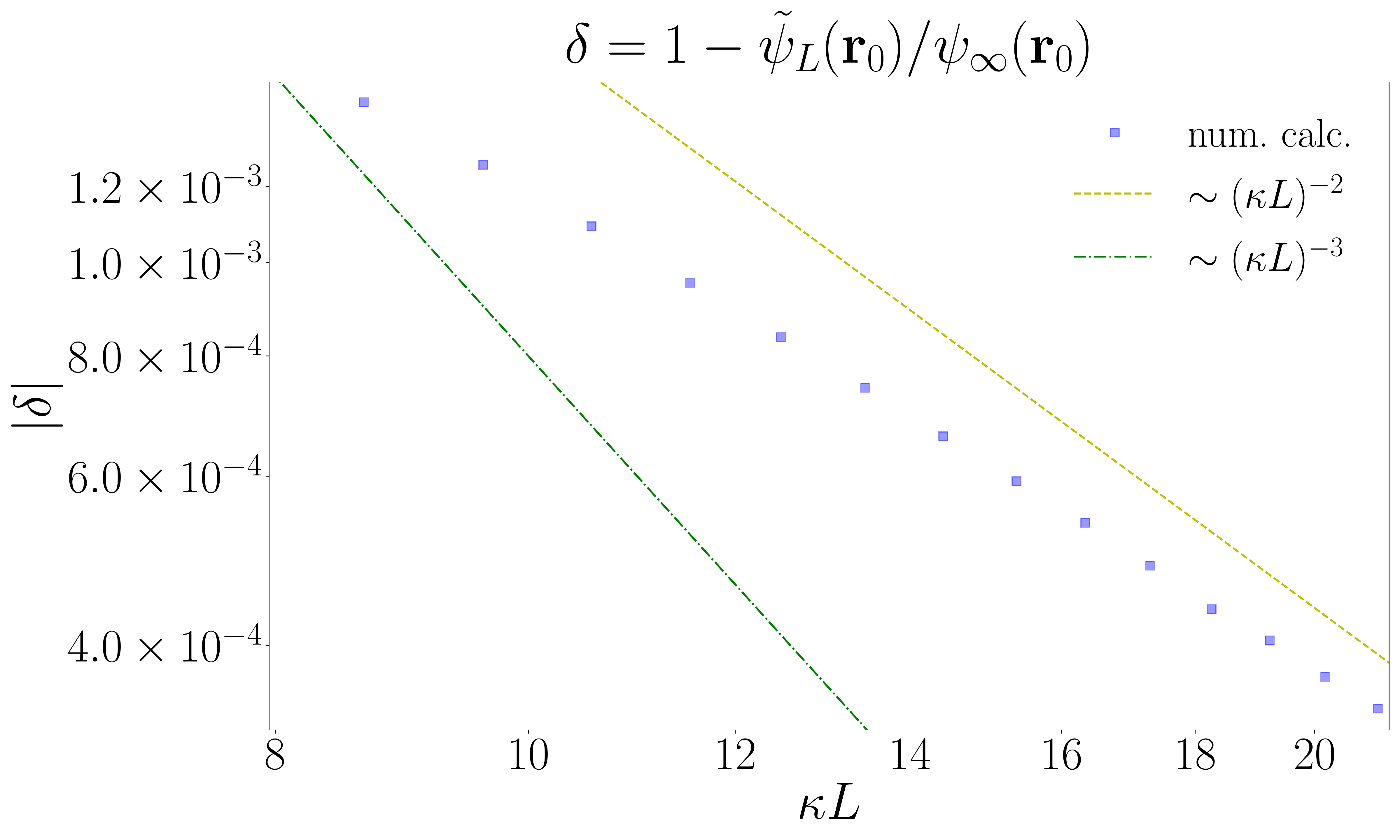}
 \caption{%
  Relative difference between $\psiinf$ and $\psip_L$, calculated as described
  in the text.
  The individual wave functions were obtained from lattice calculations using
  $a_{\text{latt}} = 1/4$.
  The dashed and dash-dotted line show functions $C_2 \times 1/(\kappa L)^2$
  and $C_3 \times 1/(\kappa L)^3$, respectively, with $C_2 = 0.175$ and $C_3 =
  0.8$.
  \label{fig:plotwave}
 }
 \end{figure}

We can now finally go back to Eq.~\eqref{eq:DeltaE-tilde-p-explicit} and
collect what we have learned.
For the first part of $\Edp'(L)$ we find:
\begin{spliteq}
 \Edp'_1(L)
 &= \frac{6}{\mu}\int_{B^2_{+z}}\dd x \dd y \frac{\partial}{\partial z}\left(
  {-}2\braket{\psiinf}{\vec r} \braket{\vec r}{\psip^{(1)}_{L,m=0}}
 \right) + \OO(\ee^{{-}(\sqrt2/2+1/2)\kappa L}) \\
 &= \frac{6}{\mu}\int_{0}^{\infty} \dd k\,\frac{1}{E_\infty+E_k}
 \sum_{l}a_{kl0}\sqrt{\frac{2l+1}{4\pi}}
 \int_{B^2_{+z}}\dd x \dd y\, \frac{\partial}{\partial z}
 \Big(\psiinf(r){\psi_{\infty,kl0}(\vec r)}\Big)
 + \OO(\ee^{{-}(\sqrt2/2+1/2)\kappa L}) \,.
\end{spliteq}
The partial derivative with respect to $z$ can be written as
\begin{equation}
 \frac{\partial}{\partial z}
 = \cos\theta \frac{\partial}{\partial r}
 + \frac{\sin\theta}{r} \frac{\partial}{\partial \theta} \,,
\label{eq:ddz}
\end{equation}
and then, noting the factor $1/r$ in front of the derivative respect to
$\theta$, we obtain, for $l \geq 1$:
\begin{spliteq}
 \abs{
  \frac{\sin\theta}{r} \frac{\partial}{\partial \theta}
  \Big(\psiinf(r){\psi_{\infty,kl0}(\vec r)}\Big)
 } &= \abs{
  \frac{\sin\theta}{r}\psiinf(r)
  \frac{\partial}{\partial \theta} \psi_{\infty,kl0}(\vec r)
 } \\
 &= \sqrt{\frac{2l+1}{4\pi}}\abs{
  \psi_{\infty,kl}(r)\psiinf(r)\frac{\sin\theta}{r}
  \frac{\partial}{\partial \theta}P_l(\cos\theta)
 } \\
 &= \sqrt{\frac{2l+1}{4\pi}}\abs{
  \psi_{\infty,kl}(r)\psiinf(r)\frac{\sin\theta}{r}
   P^1_l(\cos\theta)
 } \\
 &\leq \sqrt{\frac{2l+1}{16\pi}}\sqrt{\frac{(l+1)!}{(l-1)!}}
 \abs{\strut
  \psi_{\infty,kl}(r)\psiinf(r)
 }\times \frac{1}{r}  \,.
\end{spliteq}
In the last step above, we have used $\abs{\sin\theta} \leq 1$ and again
Eq.~(5) from Ref.~\cite{Lohoefer:1998aa} to bound $P^1_l(\cos\theta)$.
For $l = 0$, the expression vanishes trivially.
In fact, the details of the prefactors do not matter: since on
$B^2_{+z}$ we have $r\geq r_0$, it follows from the $1/r$ factor alone that the
entire contribution is $\OO\big((\kappa L)^{-1}\big) \times \psi_{\infty,kl}(r)
\psiinf(r)$.
For the contribution stemming from the first term on the left-hand side of
Eq.~\eqref{eq:ddz} we find
\begin{equation}
 \cos\theta \frac{\partial}{\partial r}
 \Big(\psiinf(r){\psi_{\infty,kl0}(\vec r)}\Big)
 = \cos\theta  P_l(\cos\theta) \sqrt{\frac{2l+1}{4\pi}}
 \frac{\partial}{\partial r}\Big(\psi_{\infty,kl}( r) \psiinf(r)\Big) \,,
\end{equation}
and we use the asymptotic forms of the scattering wave functions, given in
Eqs.~\eqref{eq:psi-scatt-asympt} and~\eqref{eq:psi-bound-asympt}, to find that
we have that
\begin{equation}
 \abs{\frac{\partial}{\partial r}\Big(\psi_{\infty,kl}(r) \psiinf(r)\Big)}
 \leq \abs{(\kappa + k) \psi_{\infty}(r) \sqrt{\frac{2}{\pi}}\frac{1}{r}}
 \times \Big(1 + \OO((\kappa L)^{-1})\Big) \,.
\end{equation}
In order to obtain this relation we have inserted the asymptotic form for
$\psiinf(r)$, performed the derivative, and then wrote the result again
as $\psiinf(r)$ plus corrections as indicated in the parentheses, similar to
what we did in order to obtain Eq.~\eqref{eq:Theta-k-L-asympt}.
The same result can be obtained using an explicit expression for the derivative
of the Whittaker function.
Then, with $\cos\theta \leq 1$ and $P_l(\cos\theta) \leq
1$, we get
\begin{equation}
 \abs{
  \frac{\partial}{\partial z}\Big(\psiinf(r){\psi_{\infty,kl0}(\vec r)}\Big)
 } \leq \sqrt{\frac{2l+1}{\pi^3}} \abs{
  \frac{(\kappa + k)}{r} \psi_{\infty}(r)
 } \times \Big(1 + \OO((\kappa L)^{-1}) \Big) \,.
\end{equation}

Together with Eq.~\eqref{eq:a_kl0_finalineq}, we arrive at
\begin{spliteq}
 \abs{\Edp'_1(L)} &\leq \frac{6}{\mu}\int_{0}^{\infty} \dd k\,
 \frac{E_\infty(\kappa + k)}{(E_\infty+E_k)^2}
 \sum_{l}{\frac{2l+1}{4\pi}}\abs{a_{kl0}}
 \left(1 + \OO\!\left(\frac{1}{\kappa L}\right)\right) \\
 & \hspace{6em} \null \times \int_{B^2_{+z}}\!\!\! \dd x \dd y\,
 \left(\psi_{\infty}(r) \sqrt{\frac{2}{\pi}}\frac{1}{r}\right)
 + \OO(\ee^{{-}(\sqrt2/2+1/2)\kappa L}) \\
 &= \OO\left(\frac{\bar\eta}{\kappa L}\right)\frac{\psiinf(r_0)}{\mu L }
 \times \int_{B^2_{+z}}\!\!\!\dd x \dd y \, \psiinf(r)
 + \OO(\ee^{{-}(\sqrt2/2+1/2)\kappa L}) \\
 &= \OO\!\left(\frac{\bar\eta}{(\kappa L)^2}\right) \times \Delta E_{0}(L) \,,
\end{spliteq}
where the sum over $l$ is performed as explained below
Eq.~\eqref{eq:psip-1-m0-sum}, and where in the last line we have proceeded as
explained below Eq.~\eqref{eq:DeltaEp-1-bound}, \ie, we have used the asymptotic
form of the Whittaker functions to simplify $\Delta E_{0}(L)$ in the limit $L
\to \infty$.
The surface integral has been performed using the asymptotic form of
$\psiinf(r)$, Eq.~\eqref{eq:psi-bound-asympt}.

Finally we can show that $\Edp'_2(L)$ is further suppressed by additional powers
of $\kappa L$.
The starting point for this is
\begin{equation}
 \Edp'_2(L) = \frac{12}{\mu}
 \int_{B^2_{+z}}\!\!\!\dd x\dd y \, \braket{\psip^{(1)}_{L}}{\vec r}
 \times \frac{\partial}{\partial z}\braket{\vec r}{\psip^{(1)}_{L}} \,.
\end{equation}
First, we can use an analysis similar to what is discussed following
Eq.~\eqref{eq:ddz} to simplify the factor involving the partial derivative
with respect to $z$:
\begin{spliteq}
 \frac{\partial}{\partial z}\braket{\vec r}{\psip^{(1)}_{L}}
 &= \sum_{k}\frac{1}{E_\infty+E_{k}}\sum_{l,{m}}
 \sqrt{\frac{2l+1}{4\pi} } a_{klm} \frac{\partial}{\partial z}
 \braket{\vec r}{\psi_{\infty,klm}} \\
 &= \sum_{k}\frac{k}{E_\infty+E_{k}}\sum_{l,{m}}
 \sqrt{\frac{2l+1}{4\pi} } a_{klm}
 \braket{\vec r}{\psi_{\infty,klm}}
 \times \left(1+\OO\left((\kappa L)^{-1}\right)\right)
 =\OO\left(\frac{\psiinf(r_0)}{(\kappa L)^{2-1/4}}\right) \,.
\end{spliteq}
That is, the partial derivative does not affect the asymptotic behavior of
$\braket{\vec r}{\psip^{(1)}_{L}}$ and we just recover the same scaling as in
Eq.~\eqref{eq:psip-m-neq-0}.
It therefore follows directly that
\begin{equation}
 \abs{\Edp'_2(L)}
 = \OO\!\left(\frac{1}{(\kappa L)^{3-1/2}}\right) \times \Delta E_{0}(L) \,,
\end{equation}
and overall we ultimately have shown that
\begin{equation}
 \Edp'(L) = \Edp'_1(L) +\Edp'_2(L)
 = \OO\!\left(\frac{\bar\eta}{(\kappa L)^2}\right) \times \Delta E_{0}(L) \,,
\end{equation}
\ie, the second correction term is ultimately suppressed compared to the
leading contribution by the same relative factor as the first correction term.

\smsection{Continuum calculations}

To assess the binding momenta and ANCs which we extract from fitting the volume
dependence, we perform infinite-volume continuum calculations for the same
systems.
These are implemented as solutions of the radial Schrödinger equation,
\begin{equation}
 {-}\frac1{2\mu}\frac{\dd^2 w(r)}{\dd r^2}
 + V(r)w(r) + V_C(r)w(r) = {-}\kappa^2 w(r) \,,
\label{eq:radseq}
\end{equation}
via the ``shooting method.''
That means that we solve the ODE~\eqref{eq:radseq} by integrating outwards from
small distances and vary $\kappa$ until the solution $w(r)$ decays to zero at
large $r$.
The $\kappa$ where this behavior is found is then the desired value for
$\kappa_\infty$.
For $d=1$, we start with $w(r_0) = 1$ and $w'(r_0) = 0$ for $r_{0} \to 0$
to find the even-parity ground state.
For $d=3$, we start with $w(r_0) = 0$, $w'(r_0) = 1$ for $r_{0} \to 0$,
which is the appropriate boundary condition for an $S$-wave ground state.
The solution for $\kappa = \kappa_\infty$ is then normalized, and the ANC is
determined by matching the tail of the normalized wave function
to the appropriate Whittaker function.
In Table~\ref{tab:V0} we list the strengths of the attractive Gaussian
potentials that we used to obtain the results reported in the main text.

\begin{table}[htbp!]
  \def\arraystretch{1.2}
  \pgfplotstabletypesetfile[
  columns = {
   gamma, v0, kappac, Ancc
  },
  columns/gamma/.style = {
   string type, column name=$\gamma$,
   column type={>{\centering\arraybackslash}p{2.5em}}
  },
  columns/v0/.style = {
   string type, column name=$V_0$,
   column type={>{\centering\arraybackslash}p{4.8em}}
  },
  columns/kappac/.style = {
   string type, column name=$\kappa_\infty$,
   column type={>{\centering\arraybackslash}p{3.8em}}
  },
  columns/Ancc/.style = {
   string type, column name=$A_\infty$, column type={c}
  },
  every head row/.style= {
   before row = {
    \TopRule
    \multicolumn{2}{c}{\ \ Parameters \ \ } &
    \multicolumn{2}{c}{\ \ Continuum result\ \ }
    \\
    \MidRule
   },
   after row = \MidDoubleRule
  },
  every row no 0/.style = {
   before row = {
    \multicolumn{4}{c}{$d=1$} \\ \MidRule
   },
  },
  every row no 3/.style = {
   before row = {
    \hline
    \multicolumn{4}{c}{$d=3$} \\ \MidRule
   },
  },
  every last row/.style = {after row = \BottomRule}
 ]{results.txt}
 \caption{%
  Detailed parameters used for the numerical results presented in the main text.
  The range of the attractive Gaussian potential is set to $R=1$ for all
  calculations.
  All quantities are given in units of the particle mass $m=1$ (see main text).
 \label{tab:V0}
 }
\end{table}

\smsection{Lattice calculation of three-nucleon system}

For the lattice calculations of the $^3$He and $^3$H systems, we take the
nucleon mass to be $m = 938.92$~MeV and set the lattice spacing to be $1.97$~fm.
We use dimensionless lattice units and our spatial volume is a periodic box with
edge length equal to $L$ lattice sites.
We denote spatial lattice sites as $\vec{n}$ and write $\hat{\vec{l}}$ for the
lattice unit vectors.
For the lattice kinetic energy operator, we use an $O(a^4)$-improved lattice
action~\cite{Lee:2008fa,Lahde:2019npb},
\begin{equation}
 K = \frac{1}{2m} \sum_{\vec n} \sum_{j} \sum_{\hat{\vec l}
 = \pm {\hat{\vec 1}},\pm {\hat{\vec 2}}, \pm {\hat{\vec 3}}}
 \left[w_0 a_j^\dagger({\vec n})a_j({\vec n})
 + w_1 a_j^\dagger({\vec n}+{\hat{\vec l}})a_j({\vec n})
 + w_2 a_j^\dagger({\vec n}+{2 \hat{\vec l}})a_j({\vec n})
 + w_3 a_j^\dagger({\vec n}+{3 \hat{\vec l}})a_j({\vec n})\right] \,,
\end{equation}
where $j$ runs over all nucleon spin and isospin indices and $w_0 = 49/36$,
$w_1 = -3/2$, $w_2 = 3/20$, $w_3 = -1/90$.
The full Hamiltonian has the form
\begin{equation}
 H = K + V_{NN} + V_{3N} + V_{C} \,.
\end{equation}
The nucleon-nucleon interaction $V_{NN}$ is an $S$-wave contact interaction
that is pointlike and independent of spin and isospin,
\begin{equation}
 V_{NN} = \frac{c_{NN}}{2} \sum_{\vec n}
 \left[\sum_j a^\dagger_j({\vec n}) a_j({\vec n}) \right]^2 \,.
\end{equation}
The three-nucleon interaction $V_{3N}$ is also an $S$-wave contact interaction
that is pointlike and independent of spin and isospin,
\begin{equation}
 V_{3N} = \frac{c_{3N}}{6} \sum_{\vec n} \left[
  \sum_j a^\dagger_j({\vec n}) a_j({\vec n})
 \right]^3 \,..
\end{equation}
For the Coulomb interaction between protons, we take
\begin{equation}
 V_{C} = \alpha \sum_{\vec n} \sum_{\vec n'} \left[
  \sum_{j = \text{protons}} a^\dagger_j({\vec n}) a_j({\vec n})
  \frac{1}{d({\vec n},{\vec n'})}
  \sum_{j' = \text{protons}} a^\dagger_{j'}({\vec n'}) a_{j'}({\vec n'})
 \right] \,,
\end{equation}
where $d({\vec n},{\vec n'})$ is the shortest distance between ${\vec n}$
and ${\vec n'}$ on the periodic lattice of length $L$.
This corresponds exactly to the truncated finite-volume Coulomb potential
discussed in the main text.
At zero separation distance we choose $d({\vec n},{\vec n})$ to equal $z_0$,
where $z_0$ is a short-distance regulator for the otherwise singular Coulomb
potential at zero distance.

The nucleon-nucleon coupling $c_{NN}$ is tuned so that the binding energy of
the dimer (deuteron) is $1.000$~MeV.
This choice is approximately halfway between the physical interaction strengths
for the 3S1-3D1 and 1S0 $NN$ channels; it corresponds to a value of
$c_{NN} = {-}0.572927$ in lattice units.
The three-nucleon coupling $c_{3N}$ is chosen so that the infinite-volume
binding energy of $^3$H is consistent with the observed value of $8.482$~MeV.
This corresponds to a value of $c_{3N} = 0.6623$ in lattice units.
The value of $z_0$, which absorbs a zero-range proton-proton coupling, is chosen
so that the infinite-volume binding energy of $^3$He is consistent with the
observed value of $7.718$~MeV.
This corresponds to a value of $z_0 = 0.7525$ in lattice units.
The calculations of the $^3$He and $^3$H finite volume spectra shown in the
main text were performed for $L$ ranging from $5$ to $18$ lattice units.

For comparison with the finite-volume extraction of ANCs, we also compute ANCs
directly from the lattice wave functions for the largest possible periodic
box length $L$.
This calculation is performed by computing the ratio
\begin{equation}
 \left(\frac{
 \bra{\psi_{{^3}{\text{He}}}}
   a^{\dagger}_{p_{\downarrow}}(\vec{n})
   a_{p_{\downarrow}}(\vec{n})
   a^{\dagger}_{p_{\uparrow}}(\vec{0})
   a_{p_{\uparrow}}(\vec{0})
   a^{\dagger}_{n_{\uparrow}}(\vec{0})
   a_{n_{\uparrow}}(\vec{0})
  \ket{\psi_{{^3}{\text{He}}}}
 }{
  \bra{\psi_{{^2}{\text{H}}}}
  a^{\dagger}_{p_{\uparrow}}(\vec{0})
  a_{p_{\uparrow}}(\vec{0})
  a^{\dagger}_{n_{\uparrow}}(\vec{0})
  a_{n_{\uparrow}}(\vec{0})
  \ket{\psi_{{^2}{\text{H}}}}
 }\right)^{1/2}.
\label{eq:wf-p-pd}
\end{equation}
For our choice of interactions, the proton-proton dimer ${^2}$He is also bound
with binding energy of $0.438$~fm in infinite volume.
We account for this when we fit the volume dependence of the \isotope[3]{He}
binding energy by using the Coulomb formula with a factor $2/3$ to account for
the asymptotic $p$-$(np)$ breakup into proton and deuteron, and the formula for
uncharged clusters with a factor $1/3$.
This is consistent with the results of Ref.~\cite{Konig:2017krd}, \ie, we
include all relevant $1+2$ breakup channels with proper combinatorial factors.
We explicitly account for the small difference in the binding momenta for the
two channels in infinite volume, and we assume that the ANCs are the same for
both channels, so that the fit still only needs to determine two parameters.
Moreover, we can extract the wave function for this channel by computing the
ratio
\begin{equation}
 \left(\frac{
   \bra{\psi_{{^3}{\text{He}}}}
   a^{\dagger}_{n_{\uparrow}}(\vec{n})
   a_{n_{\uparrow}}(\vec{n})
   a^{\dagger}_{p_{\uparrow}}(\vec{0})
   a_{p_{\uparrow}}(\vec{0})
   a^{\dagger}_{p_{\downarrow}}(\vec{0})
   a_{p_{\downarrow}}(\vec{0})
   \ket{\psi_{{^3}{\text{He}}}}
  }{
   \bra{\psi_{{^2}{\text{He}}}}
   a^{\dagger}_{p_{\uparrow}}(\vec{0})
   a_{p_{\uparrow}}(\vec{0})
   a^{\dagger}_{p_{\downarrow}}(\vec{0})
   a_{p_{\downarrow}}(\vec{0})
   \ket{\psi_{{^2}{\text{He}}}
 }}\right)^{1/2} .
\label{eq:wf-n-pp}
\end{equation}

We apply the formulas above along lattice axis $\hat{\vec 1}$.
The ANC is then extracted from the constant of proportionality with the corresponding
asymptotic form of the wave function for large $r = |\vec{n}|$.
Specifically, we use the ansatz $\psi_{L,0}(r)$, defined in the main text
between Eqs.~(10) and~(11), to approximate the finite-volume
wave function:
\begin{equation}
 \psi_{L}(r) \approx \psi_{L,0}(r) \approx \psiinf(r) + \psiinf(L-r) \,,
\end{equation}
with the fit range restrict to $r\in(0+\epsilon,L-\epsilon)$ for some $\epsilon >0$, so that the asymptotic form $\psiinf(r) = A_\infty {W_{{-}\bar\eta,1/2}(2\kappa r)}/{(\sqrt{4\pi} r)}$ applies.
From Eq.~\eqref{eq:delta-tildaphi-0}, we  find that $\psi_{L,0}(r)$ agrees
with $\psi_{L}(r)$ up to a power-law correction.
The largest discrepancy emerges at the boundary of the box, and
\begin{equation}
 \psi_{L,0}(L/2) = \psi_L(L/2)
 \times \left[1+\OO\left(\frac{\bar\eta}{(\kappa L)^2}\right)\right] \,.
\end{equation}
We choose $L$ as large as possible to suppress this correction.
In practice, we fit the logarithm of the ratios in Eqs.~\eqref{eq:wf-p-pd}
and~\eqref{eq:wf-n-pp} to the logarithm of the approximate $\psi_{L,0}(r)$.
This fit determines both $\kappa$ and the ANC.
To determine the uncertainty estimates reported in the main text, we vary
$\epsilon$ and calculate the mean and mean deviation of the fitted ANCs.

\smsection{Finite-volume Coulomb potential via zero-mode subtraction}
\label{sec:ZeroModeSubtraction}

An alternative approach to defining the Coulomb interaction in a periodic box,
introduced in Ref.~\cite{Beane:2014qha}, is to remove the ``zero mode'' that
enters in the momentum-space representation of the potential.
Recalling that we use $\gamma = 2\mu \alpha Z_1 Z_2$ to parameterize to strength
of the Coulomb potential for a system with reduced mass $\mu$, this approach
leads to defining
\begin{equation}
 U_L(\vec{r}) = \frac{\gamma}{2\mu\pi L} \sum_{\vec{n} \neq \vZero}
 \frac{1}{\vec{n}^2}\ee^{\frac{\ii 2\pi \vec{n} \cdot \vec{r}}{L}} \,.
\label{eq:FVQED}
\end{equation}
Following Ref.~\cite{Luscher:1990ux}, the free Green's function in a cubic
periodic box can be written as
\begin{equation}
 G(\vec r,k^2) = \frac{1}{4\pi^2} \sum_{\vec{n} \in \mathbb{Z}^3}
 \frac{1}{\vec{n}^2-q^2}\ee^{\frac{\ii 2\pi \vec{n} \cdot \vec{r}}{L}}
 \mathtext{,} q = \frac{kL}{2\pi} \,,
\label{eq:G-latt-0}
\end{equation}
and by comparison with Eq.~\eqref{eq:FVQED} we find that
\begin{equation}
 U_L(\vec{r}) = \frac{2\pi\gamma}{\mu} \lim_{k\to 0} \left[
  G(\vec{r}, k^2) + \frac{1}{\pi L} \frac{1}{q^2}
 \right] \,.
\end{equation}
This is useful because it allows us to study the behavior of $U_L(\vec r)$ based
on known properties of the Green's function.
Specifically, one can show that for $\vec r \in [-L/2,L/2]^3$ it holds that
\begin{equation}
 G(\vec{r},k^2) = \frac{\cos (kr)}{4\pi r}
 + \sum_{l=0}^{\infty}\sum_{m={-}l}^l g_{lm} Y_{lm}(\theta,\phi) j_l(k r) \,,
\label{eq:G-latt-expanded}
\end{equation}
where the coefficients $g_{lm}$ are given in terms of a zeta function as
\begin{equation}
 g_{lm} = \frac{\ii^l}{\pi L q^l} \mathcal{Z}_{lm}(1;q^2) \,,
\end{equation}
see Ref.~\cite{Luscher:1990ux} for details.
The spherical harmonics $Y_{lm}(\theta,\phi)$ in Eq.~\eqref{eq:G-latt-expanded}
capture the spatial orientation of $\vec{r}$.
To analyze the limit $k\to0$, we can make use of the known behavior of the
spherical Bessel functions for small argument:
\begin{equation}
 j_l(k r) \to \frac{(kr)^l}{(2l+1)!!} \mathtext{for} kr \to 0 \,.
\label{eq:Bessel-kr-0}
\end{equation}
Using this and inserting the definition of the $g_{lm}$, we obtain
\begin{equation}
 \lim_{k\to 0} G(\vec{r},k^2) = \frac{1}{4\pi r}
 + \frac{1}{\pi L} \sum_{l,m} (2\pi \ii)^l \, Y_{lm}(\theta,\phi)
 \times \left(\frac{r}{L}\right)^l
 \times \lim_{q\to 0} \mathcal{Z}_{lm}(1;q^2) \,.
\label{eq:G-latt-limit}
\end{equation}
Cubic symmetry limits the combinations of $l$ and $m$ that can appear in the
sum.
The first two allowed terms involve $\mathcal{Z}_{00}$ and $\mathcal{Z}_{40}$,
so that we find
\begin{equation}
 \lim_{k\to 0} G(\vec{r},k^2)
 = \frac{1}{4\pi r} + \frac{1}{\pi L} Y_{00}(\theta,\phi)
 \times \lim_{q\to 0} \mathcal{Z}_{00}(1;q^2)
 + \OO\left(\frac{r^4}{L^5}\right) \,,
\label{eq:G-latt-zeta}
\end{equation}
using Eq.~\eqref{eq:Bessel-kr-0}.
The remaining zeta function can furthermore be simplified with the help of
Eq.~(6.20) in Ref.~\cite{Luscher:1990ux}, which leads to
\begin{equation}
 \sqrt{4\pi} \mathcal{Z}_{00}(1;q^2)
 = {-}\frac{1}{q^2} + C_0 + \OO(q^2) \,,
\end{equation}
with a numerical constant given by $C_0/\pi \approx {-}2.83$.
Inserting this into Eq.~\eqref{eq:G-latt-zeta} gives
\begin{equation}
 \lim_{k\to 0} G(\vec{r},k^2)
 = \lim_{q\to 0} \frac{1}{4\pi}\left[
  \frac{1}{r} + \frac{C_0}{\pi L} - \frac{1}{q^2}
  +\OO\left(\frac{r^4}{L^5}\right)
 \right] \,,
\end{equation}
and, noting that the singular term cancels out, we finally arrive at
\begin{equation}
 U_L(\vec{r}) = \frac{\gamma}{2\mu r} + \frac{\gamma C_0}{2\mu\pi L}
 + \OO\left(\frac{r^4}{L^5}\right) \,.
\label{eq:FVQED-expanded}
\end{equation}

Equation~\eqref{eq:FVQED-expanded} is valid within the central box, and by
construction $U_L(\vec{r})$ is cubic periodic.
With that observation it is clear that the first term in
Eq.~\eqref{eq:FVQED-expanded} corresponds exactly to the truncated periodic
Coulomb potential $V_{C,\{L\}}(\vec{r})$ used in our approach.
The derivation above shows that the leading correction departing from that
simple form is a constant overall shift (without dependence on $\vec{r}$)
that carries over directly to the energy spectrum and can easily be corrected
for.
The next term, of order $r^4/L^5$, however, is not particularly suppressed for
$r$ near the box boundary, and it furthermore depends on the direction of
$\vec{r}$.

In fact, since this corrections scales like a positive power of $r$, it does
not correspond to a short-range potential and therefore it can give rise to
contribution to the binding energy that are not exponentially suppressed for
large $L$.
We note that for a limited regime where $\kappa$ and $\gamma$ are sufficiently
small for the leading volume dependence of exponential order to dominate over
contributions from the higher-order terms in Eq.~\eqref{eq:FVQED-expanded}.
In this case, it may still be
possible to extract approximate ANCs from calculations involving $U_L$
by explicitly correcting for the leading constant shift $\gamma C_0/(2\mu\pi L)$.
In practice, however, the complicated behavior of $U_L$ as a function of $L$
makes this approach less useful for extracting infinite-volume asymptotic
properties of bound states.
The primary application of our main result therefore concerns Lattice EFT
calculations, where $V_{C,\{L\}}(\vec{r})$ is a simple and natural choice
for implementing the Coulomb potential.

\end{widetext}

\end{document}